\documentclass[epj,nopacs]{svjour}

\def\makeheadbox{}
\usepackage{lmodern}
\usepackage[T1]{fontenc}

\usepackage{tikz}

\usetikzlibrary{shapes.geometric, arrows,positioning,fit,scopes,backgrounds}
\tikzstyle{box} = [rectangle, text width=4cm, minimum height=1cm, text centered, draw=black, fill=red!30, thick, inner sep=5pt]
\tikzstyle{arrow} = [very thick,->,>=stealth]
\tikzstyle{depbox} = [rectangle, text width=3.1cm, minimum height=1cm, text centered, draw=black, fill=green!30, thick, inner sep=3.5pt]
\tikzstyle{pt} = [draw=black, thick, fill=blue!75!black!25, inner sep=8pt]

\DeclareMathSymbol{\Gamma}{\mathalpha}{operators}{0}
\DeclareMathSymbol{\Delta}{\mathalpha}{operators}{1}
\DeclareMathSymbol{\Theta}{\mathalpha}{operators}{2}
\DeclareMathSymbol{\Lambda}{\mathalpha}{operators}{3}
\DeclareMathSymbol{\Xi}{\mathalpha}{operators}{4}
\DeclareMathSymbol{\Pi}{\mathalpha}{operators}{5}
\DeclareMathSymbol{\Sigma}{\mathalpha}{operators}{6}
\DeclareMathSymbol{\Upsilon}{\mathalpha}{operators}{7}
\DeclareMathSymbol{\Phi}{\mathalpha}{operators}{8}
\DeclareMathSymbol{\Psi}{\mathalpha}{operators}{9}
\DeclareMathSymbol{\Omega}{\mathalpha}{operators}{10}

\renewcommand{\Re}{\text{Re}}
\renewcommand{\Im}{\text{Im}}

\newcommand{\ztwo}{\mathbb{Z}_2}

\usepackage{slashed}
\usepackage{graphics}
\usepackage[numbers,sort&compress]{natbib}
\usepackage{mathtools}
\usepackage{braket}
\usepackage{xspace}
\usepackage{xcolor}
\usepackage{tabularx, booktabs}
\usepackage{amsmath}
\usepackage{amssymb}

\usepackage{hyperref}
\hypersetup{colorlinks=true,linkcolor=teal,citecolor=teal}
\usepackage{cleveref}
\usepackage{xurl}

\newcommand{\gev}{\ensuremath{\,\text{GeV}}\xspace}

\newcommand{\msbar}{\ensuremath{\overline{\text{MS}}}\xspace}
\newcommand{\pt}{\texttt{PhaseTracer}\xspace}
\newcommand{\PT}{\texttt{PhaseTracer}\xspace}

\makeatletter
\g@addto@macro\bfseries{\boldmath}
\makeatother

\usepackage[scale=0.8]{GoMono}

\definecolor{keyword}{HTML}{008000}
\definecolor{emph}{HTML}{0000FF}
\definecolor{string}{HTML}{A52A2A}
\definecolor{comment}{rgb}{0.0, 0.44, 1.0}
\definecolor{back}{HTML}{F8F8F8}
\definecolor{arrow}{HTML}{745334}

\usepackage{listings, dirtree, mdframed}

\newcommand{\code}{\texttt}

\usepackage{microtype}
\usepackage[english]{babel}
\newcommand{\origunderscore}{}
\let\origunderscore\_
\renewcommand{\_}{\allowbreak\origunderscore}

\usepackage[htt]{hyphenat}
\begin{document}

\title{\code{PhaseTracer2}: from the effective potential to gravitational waves}

\newcommand{\addemail}[1]{\url{#1}}

\mail{\addemail{peter.athron@njnu.edu.cn},  \addemail{csaba.balazs@monash.edu}, \addemail{andrew.fowlie@xjtlu.edu.cn}, \addemail{lachlan.morris@monash.edu}, \addemail{william.searle@monash.edu.au}, \addemail{xiaoyang@itp.ac.cn}, \addemail{yang.phy@foxmail.com}}

\author{Peter Athron\inst{2,3}
\and Csaba Bal\'azs\inst{4}
\and Andrew Fowlie\inst{5}
\and Lachlan Morris\inst{4}
\and William Searle\inst{4} 
\and Yang Xiao\inst{6,7}
\and Yang Zhang\inst{1,8}
}

\titlerunning{from the effective potential to gravitational waves}
\authorrunning{\code{PhaseTracer2}}

\institute{School of Physics, Henan Normal University, Xinxiang 453007, P. R. China
\and Institute of Theoretical Physics and Institute of Physics Frontiers and Interdisciplinary Sciences, Nanjing Normal University, Wenyuan Road, Nanjing, Jiangsu, 210023, China
\and Ministry of Education Key Laboratory of NSLSCS, Nanjing Normal University, Nanjing 210023, China
\and School of Physics and Astronomy, Monash University, Melbourne 3800 Victoria, Australia
\and Department of Physics, School of Mathematics and Physics, Xi’an
Jiaotong-Liverpool University, Suzhou 215123, China
\and CAS Key Laboratory of Theoretical Physics, Institute of Theoretical Physics, Chinese Academy of Sciences, Beijing 100190, China
\and School of Physical Sciences, University of Chinese Academy of Sciences,  Beijing 100049, China
\and School of Physics, Zhengzhou University, Zhengzhou 450001, China
}

\date{}
%
\abstract{
In recent years, the prospect of detecting gravitational waves sourced from a strongly first-order cosmological phase transition has emerged as one of the most exciting frontiers of gravitational wave astronomy. Cosmological phase transitions are an essential ingredient in the Standard Model of particle cosmology, and help explain the mechanism for creation of matter in the early Universe, provide insights into fundamental theories of physics, and shed light on the nature of dark matter. This underscores the significance of developing robust end-to-end tools for determining the resulting gravitational waves from these phase transitions. In this article we present \code{PhaseTracer2}, an improved version of the \code{C++} software package \code{PhaseTracer}, designed for mapping cosmological phases and transitions in Standard Model extensions of multiple scalar fields. Building on the robust framework of its predecessor, \code{PhaseTracer2} extends its capabilities by including new features crucial for a more comprehensive analysis of cosmological phase transitions. It can calculate more complex properties, such as the bounce action through the path deformation method or an interface with \code{BubbleProfiler}, thermodynamic parameters, and gravitational wave spectra. Its applicability has also been broadened via incorporating the dimensionally reduced effective potential for models obtained from \code{DRalgo}, as well as calculations in the \msbar and OS-like renormalisation schemes. This modular, flexible, and practical upgrade retains the speed and stability of the original \code{PhaseTracer}, while significantly expanding its utility.
} 
\maketitle
\tableofcontents 

\section{Introduction}\label{sec:Intro}

The detection of gravitational waves created during cosmological first-order phase transitions would open a new frontier in both particle physics and cosmology~\cite{Athron:2023xlk}.  First-order phase transitions, as predicted in many extensions of the Standard Model, play a crucial role in the early Universe.  They can inform us about symmetries that preceded those of the Standard Model, about the mechanism of generating the cosmic matter-antimatter asymmetry, or the observed abundance of dark matter.  This direct insight into the physics of the early Universe is possible because these transitions generate a stochastic gravitational wave background that could be observed by upcoming experiments such as LISA~\cite{amaro2017laser}, Taiji~\cite{10.1093/nsr/nwx116}, and Tianqin~\cite{TianQin:2015yph}.

Modelling these phase transitions requires computational tools that account for complex thermal corrections and the dynamic evolution of scalar fields across cosmological temperatures. 
\code{CosmoTransitions}~\cite{wainwright2012cosmotransitions} is a long-standing and widely utilized tool designed to output phases, critical temperatures and nucleation temperatures. Developed in \code{Python}, \code{CosmoTransitions} offers ease of use, albeit with slower performance in high-dimensional field spaces. It defaults to taking the potential using the modified minimal subtraction \msbar scheme as input and does not incorporate specific daisy corrections. Meanwhile, it addresses symmetry issues by simply discarding parts of the field space. \code{BSMPT}~\cite{Basler_2019,Basler:2020nrq,basler2024bsmptv3toolphase} is a \code{C++} software that is currently being maintained and expanded. \code{BSMPT1} employs the bisection method to calculate the critical temperature, from which it derives the strength of the electroweak phase transition. It utilizes the `on-shell' renormalization scheme and incorporates daisy corrections. The main upgrade in \code{BSMPT2} is the implementation of the computation for the baryon asymmetry of the Universe. \code{BSMPT3} has been extended to track the phases at various temperatures and compute the transition probabilities between them. Consequently, it is capable of determining nucleation, percolation, and completion temperatures, as well as estimating the stochastic gravitational wave background generated by the phase transition.

\code{PhaseTracer2} builds on the framework of its predecessor~\cite{Athron:2020sbe}, offering significant improvements in both the computational algorithms and the theoretical models used to describe the effective potential of scalar fields.  These include methods for computing the bounce action via path deformation or \code{BubbleProfiler}~\cite{Athron_2019}, various thermodynamic parameters of the transition, and the gravitational wave spectra, all of which are essential for describing phase transitions and their gravitational remnants.  

Importantly, via its extensive \code{EffectivePotential} library, \code{PhaseTracer2} incorporates predefined effective potentials for several beyond the Standard Model theories.  These effective potentials are given in four dimensions, in various renormalization schemes such as the \msbar and on-shell (OS)-like schemes, and dimensionally reduced three dimensional form \cite{Ginsparg_1980, Appelquist_Pisarski_1981, Braaten_Nieto_1995, Farakos_Kajantie_Rummukainen_Shaposhnikov_1994,Kajantie:1995dw,Nadkarni:1982kb}.  This enables analysis using a variety of methods, including the high-temperature expansion and the $\hbar$ expansion, which are critical for understanding different aspects of the phase transition.
A key component of the \code{PhaseTracer2} improvements are effective potentials calculated in the dimensional reduction framework, which simplifies the high-energy 4D theory into an effective 3D theory, providing a tractable framework for studying the low-energy dynamics of the phase transition \cite{Ekstedt_2023}.  This 3D effective field theory (3DEFT) is valid in regimes where thermal masses dominate, and serves as a powerful tool for capturing the leading thermal corrections in the high-temperature limit.\footnote {The validity of 3DEFT is restricted to certain energy regimes so its applicability is model and scenario dependent. In cases where it breaks down, alternative thermal resummation approaches, such as the optimised partial dressing \cite{Curtin:2016urg, Curtin:2022ovx}, may be able to help.} Recent studies have shown that the results of 2-loop 3DEFT agree well with non-perturbative lattice simulations and are insensitive to both the renormalization scale and gauge parameters~\cite{Niemi:2024vzw,Niemi:2024axp,Croon:2020cgk}. This implies that 3DEFT can provide a more accurate effective potential, leading to more precise thermodynamic properties. Therefore, extending the 3DEFT module in \code{PhaseTracer} is a crucial step toward achieving accurate calculations of the phase transition. 

By separating computational methods from the physical models themselves, \code{PhaseTracer2} maintains flexibility, allowing users to apply various computational approaches to a wide range of theoretical scenarios.  The tool is particularly suited for probing the detailed properties of phase transitions, such as transition strength, nucleation rate and gravitational wave signals, in extensions of the Standard Model that contain multiple scalar fields.  

The rest of the paper is structured as follows. In \cref{sec:QuickStart}, we provide quick-start instructions for the installation and running of \code{PhaseTracer2}. 
We describe the new features added in the upgraded version of \code{PhaseTracer2} beginning in \cref{sec:EffPot}, where we detail the effective potential and describe the new model implementations we now include. This proceeds a discussion of the bounce action calculator found in \cref{sec:TransProp}, where we also describe the calculation of observables such as the nucleation rate and phase transition duration. \Cref{sec:GravWaves} concludes the physics discussion with a summary of the methods we employ for calculating gravitational wave spectra. \Cref{sec:UserGuide} then provides a description of the \code{PhaseTracer2} code, detailing each class including relevant members and methods, as well as example use cases of this functionality. This leads to \cref{sec:Examples}, in which we provide a set of examples that exhibits the new capacity of our upgraded code. Lastly, we conclude in \cref{sec:Conc}.

\section{Quick start}\label{sec:QuickStart}

\subsection{Requirements}

The library requirements for constructing \code{PhaseTracer2} are similar to those of \code{PhaseTracer}:
\begin{itemize}
    \item A \code{C++14} compatible compiler
    \item \href{https://nlopt.readthedocs.io/en/latest/}{\code{CMake}}, version 2.8.12 or higher.
    \item The \href{https://nlopt.readthedocs.io/en/latest/}{\code{NLopt}} library, version 2.4.1 or higher.
    \item The \href{https://eigen.tuxfamily.org/index.php?title=Main_Page}{\code{Eigen}} library, version 3.1.0 or higher.
    \item \href{https://www.boost.org/}{\code{Boost}} libraries, version 1.53.0 or higher, specifically:
    \begin{itemize}
        \item \code{Boost.Filesystem}
        \item \code{Boost.Log}
        \item \code{Boost.Program\_options}
    \end{itemize}
    \item The \href{https://www.alglib.net/}{\code{AlgLib}} library, version 4.02.0 or higher
    \item The \href{https://www.gnu.org/software/gsl/}{\code{GLS}} library, version 2.8 or higher
\end{itemize}
On Ubuntu or Debian based distributions, these can be installed using:
\begin{lstlisting}[language=bash]
$ sudo apt install libalglib-dev libnlopt-cxx-dev libeigen3-dev libboost-filesystem-dev libboost-log-dev libboost-program-options-dev libgsl-dev  
\end{lstlisting}
On Fedora,
\begin{lstlisting}[language=bash]
$ sudo dnf install alglib-devel nlopt-devel eigen3-devel boost-devel gsl-devel
\end{lstlisting}

To visualise results calculated by \code{PhaseTracer2}, the package utilises plotting scripts that require the following:
\begin{itemize}
    \item Python, and the external modules \code{scipy}, \code{numpy}, \\ \code{matplotlib}, and \code{pandas}
    \item {gnuplot}
\end{itemize}
On Ubuntu or Debian based distributions, these can be installed using:
\begin{lstlisting}[language=bash]
$ sudo apt install gnuplot python3-numpy python3-scipy python3-matplotlib python3-pandas
\end{lstlisting}
For Fedora, the package names are identical in \code{dnf}.

In addition to the above prerequisites, \code{PhaseTracer2} features models from both \code{FlexibleSUSY} and \code{BSMPT}, and it also has the capability to utilize \code{BubbleProfiler} for the purpose of computing bounce actions. These come with additional dependencies as documented in their respective manuals; see \cite{ATHRON2018145, ATHRON2015139} for information regarding \code{FlexibleSUSY}, \cite{Basler_2019, Basler:2020nrq, basler2024bsmptv3toolphase} for information regarding \code{BSMPT}, \cite{Athron_2019} for information regarding \code{BSMPT}.

\subsection{Download and installation}

\code{PhaseTracer2} can be obtained by cloning
\begin{lstlisting}[language=bash]
$ git clone https://github.com/PhaseTracer/PhaseTracer
\end{lstlisting}
or downloading \url{https://github.com/PhaseTracer/PhaseTracer/archive/refs/heads/master.zip}. To build \code{PhaseTracer2} on a UNIX-based system, we utilise the standard \code{cmake} build system by executing the following:
\begin{lstlisting}[language=bash]
$ cd PhaseTracer
$ mkdir build
$ cd build
$ cmake ..
$ make
\end{lstlisting}
This will compile all examples. If the user wishes to compile a particular example, it can be done by running:
\begin{lstlisting}[language=bash]
$ make <run_model_name>
\end{lstlisting}
If the user desires to build with \code{BubbleProfiler} for the action calculation, they can instead run:
\begin{lstlisting}[language=bash]
$ cmake -D BUILD_WITH_BP=ON ..
\end{lstlisting}
During the execution of this command, \code{BubbleProfiler} is installed within the main package directory.
Similarly, \code{FlexibleSUSY} may be included by
\begin{lstlisting}[language=bash]
$ cmake -D BUILD_WITH_FS=ON ..
\end{lstlisting}
Both \code{FlexibleSUSY} and \code{BSMPT} are disabled by default.

The executables built from the examples shipped with \code{PhaseTracer2} can be found within the \code{bin/} sub-directory of the main package directory, and can be run by executing
\begin{lstlisting}[language=bash]
$ ./bin/<run_model_name>
\end{lstlisting}
for the various models implemented. For a comprehensive list, please refer to \cref{sec:Examples}, or to the original \code{PhaseTracer} manual~\cite{Athron:2020sbe}. As an example, the default one-dimensional test model can be run using:
\begin{lstlisting}[language=bash]
$ ./bin/run_1d_test_model -d
\end{lstlisting}
Here the \code{d} flag specifies that we wish to provide debugging information, including plots.

\subsection{Extending with \code{TransitionSolver}}
\label{sec:TS}
\PT may also be run within \code{TransitionSolver} to find the true cosmological phase history and obtain additional properties of the phase transition. We simultaneously release the beta version 0.1 here, with proper documentation to follow.  To use \code{TransitionSolver}, after installing \PT, download via git: 
\begin{lstlisting}[language=bash]
$ git clone https://github.com/TransitionSolver/TransitionSolver 
\end{lstlisting}
The current version of \code{TransitionSolver} is not documented and may not work for all cases.  \code{TransitionSolver} has been developed primarily with electroweak phase transitions in mind though it is hoped it can be used in other phase transitions too, but this may currently require some testing, adaption and understanding of what is needed from the user. \code{TransitionSolver} was first developed for Ref.~\cite{Athron:2022mmm} where details on how the checks for completion are provided. \code{TransitionSolver} has also been used in Refs.~\cite{Athron:2023mer,Athron:2023rfq} where some further details on the calculations may be found.   So far it has been used in a toy model of supercooling, a beyond Standard Model (BSM) model with a non-linear realisation of the Higgs mechanism~\cite{Kobakhidze:2016mch} and the scalar singlet extension of the standard model.  Very basic setup and usage instructions may be found in the README file in \code{TransitionSolver}, but more detailed documentation will not be provided until the manual \cite{Athron:2024tbc} and v1.0 are released.  If you use \code{TransitionSolver} before the full manual is released, please cite Ref.~\cite{Athron:2022mmm}.



\section{The Effective Potential}\label{sec:EffPot}
The effective potential is an essential input to any analysis of phase transitions. However, there are many subtle issues related to the use and evaluation of the effective potential, 
with an active literature exploring state-of-art approaches to evaluating the effective potential which include the use of dimensional reduction \cite{Ginsparg_1980, Appelquist_Pisarski_1981, Farakos_Kajantie_Rummukainen_Shaposhnikov_1994, Kajantie_Laine_Rummukainen_Shaposhnikov_1996, Braaten_Nieto_1995} or the optimised partial dressing improvement \cite{Curtin:2016urg,Curtin:2022ovx} to daisy resummation. At the same time the form of the effective potential also depends on many choices, including the gauge and renormalisation scheme used, while in some studies simpler forms of the effective potential, e.g.\ using the high temperature expansion, may be deemed most suitable. Each unique combination of these choices can result in not only minor quantitative differences in the potential, but even entirely new qualitative behaviour \cite{athron2022arbitrary, Croon_Gould_Schicho_Tenkanen_White_2021}. %

In \pt the effective potential for new models is created 
by the user implementing a class which derives from either the \code{Potential} or \code{OneLoopPotential} classes.  The latter provides methods to automate the addition of higher-order corrections once model specific details, such as the field dependent masses, are provided in the derived user-written class.  While this makes the process of adding the effective potential for new models much easier and less error prone, in the first release of \pt this had the major disadvantages that only limited forms of one-loop potentials were supported.

In \code{PhaseTracer2} we have made important adjustments to the effective potential classes, which significantly extends the possible forms of the effective potential that can be used while still taking advantage of the automation of the \code{OneLoopPotential} class, as well as changes that makes it possible to add potentials for new models that are gauge invariant or use state-of-the-art dimensional reduction.  In this section, we review the various forms of the effective potential that can now be selected within the \code{PhaseTracer2} framework.  We point the user to the key methods that enable this and to concrete examples distributed with \code{PhaseTracer2} illustrating how. 

\subsection{One-loop corrections}

\subsubsection{\texorpdfstring{\msbar}{MS-bar} and \texorpdfstring{$R_{\xi}$}{Rxi} Gauge} \label{sub:MSandRxi}

Using the $\msbar$ renormalization scheme and $R_{\xi}$ gauge is a standard approach to the effective potential, utilising the Coleman-Weinberg method \cite{Coleman_Weinberg_1973}. The potential is written as a perturbative loop expansion, which to one-loop order is given by:
\begin{equation}
    V(\phi, T) = V_0(\phi) + V_{\text{CW}}(\phi) + V_{T}(\phi, T).
\end{equation}
In the standard $R_{\xi}$ gauge, with gauge-fixing term \cite{Rxi_gauge_1972}\footnote{Note this expression distinguishes between the dynamical field $\phi$, and the background field (VEV), $\phi_c$.}
\begin{equation}
    \mathcal{L}_{\text{gf}} = - \frac{(\mathcal{F}^a)^2}{2 \xi}; \ \mathcal{F}^a = \partial_{\mu}A^{\mu, a} - \xi \phi_i (g T^a \phi_c)_i,
\end{equation}
and the \msbar scheme, the one-loop zero-temperature contribution has the following form \cite{Patel_2011}:
\begin{equation}
    \begin{split}\label{eq:MSOneLoopZeroTemp}
    V_{\text{CW}}(\phi) ={} & \frac{1}{4 (4 \pi)^2} \bigg [ \sum_{i \in h} n_i m_i^4(\phi, \xi) \bigg ( \log \bigg( \frac{m_i^2(\phi, \xi)}{Q^2}\bigg) - \frac{3}{2} \bigg ) \\
   & + \sum_{a \in V} n_a m_a^4(\phi) \bigg ( \log \bigg( \frac{m_a^2(\phi)}{Q^2}\bigg) - \frac{5}{6} \bigg ) \\
   & - \sum_{G \in \text{FP}} m_G^4(\phi, \xi) \bigg ( \log \bigg( \frac{m_G^2(\phi, \xi)}{Q^2}\bigg) - \frac{3}{2} \bigg ) \\
   & -  \sum_{I \in f} n_I m_I^4(\phi) \bigg ( \log \bigg( \frac{m_I^2(\phi)}{Q^2}\bigg) - \frac{3}{2} \bigg ) \bigg],
    \end{split}
\end{equation}
where $Q$ represents the renormalisation scale and $m_i$ are the field-dependent \msbar mass eigenstates which are implicitly RG scale dependent. Summation over $i \in h$ represents scalar fields, which can be separated into both physical Higgs and Goldstone boson contributions. Here $m_i^2(\phi, \xi)$ are not eigenvalues of the Hessian, but instead are eigenvalues of 
\begin{equation}
    M^2(\phi) + \xi m_A^2(\phi), 
\end{equation}
in which $M^2(\phi)$ \textit{is} the conventional Hessian: 
\begin{equation}\label{eq: scalar mass}
    M^2(\phi)_{ij} = \frac{\partial^2 V_0}{\partial \phi_i \partial \phi_j}.
\end{equation}
The masses,
\begin{equation}\label{eq:mAij}
    m_A^2(\phi)_{ij} = (g T^a \phi)_i (g T^a \phi)_j    
\end{equation}
arise due to the gauge fixing term. Summation over $a \in V$ represents both transverse and longitudinal massive gauge boson contributions, whose masses $m_a^2(\phi)$ are the eigenvalues of the gauge-boson mass matrix:
\begin{equation}\label{eq:mVab}
    m_V^2(\phi)^{ab} = (g T^a \phi)_i (g T^b \phi)_i.
\end{equation}
Note the distinction between $m_A^2(\phi)_{ij}$ and $m_V^2(\phi)^{ab}$ in \cref{eq:mAij,eq:mVab}. As they are built from the same object, $g T^a \phi_i$, they share the same eigenvalues \cite{Patel_2011}. Summation over $G \in \text{FP}$ represents Faddeev-Popov ghost contributions needed in the $R_{\xi}$ gauge, for which there is a contribution corresponding to each vector boson. The mass of these ghost terms are related to their corresponding vector boson term via:
\begin{equation}\label{eq:Faddeev-Popov}
   m_{G}^2(\phi, \xi) = \xi m_{\text{V}}^2(\phi),
\end{equation}
such that these terms only contribute in the $\xi \ne 0$ gauges. Finally, summation over $I \in f$ represents all (Weyl) fermion contributions. The associated degrees of freedom are $n_i = 1$, $n_a = 3$, and $n_I = 2$.

Prior to resummation, all temperature dependence of the effective potential at one loop order is contained within the following correction \cite{Patel_2011}:
\begin{equation}
    \begin{split}
        V_T(\phi, T) = \frac{T^4}{2 \pi^2} \bigg [ & \sum_{i \in h} n_i J_B \bigg ( \frac{m_i^2(\phi, \xi)}{T^2} \bigg ) \\
        & + \sum_{a \in V} n_a J_B \bigg ( \frac{m_a^2(\phi)}{T^2} \bigg ) \\
        & - \sum_{G \in \text{FP}} n_G J_B \bigg ( \frac{m_G^2(\phi, \xi)}{T^2} \bigg ) \\
        & - \sum_{I \in f} n_I J_F \bigg ( \frac{m_I^2(\phi)}{T^2} \bigg ) \bigg ],
        \label{eq:oneloopFiniteTemp}
    \end{split}
\end{equation}
where $J_B(z^2)$ and $J_F(z^2)$ are the bosonic and fermionic thermal functions:
\begin{equation}
    J_{B/F}(z^2) = \int_0^{\infty} dx \ x^2 \ln (1 \pm e^{- \sqrt{x^2 + z^2}} ),
\end{equation}
with the negative sign corresponding to the bosonic function.

In \code{PhaseTracer2}, the gauge parameter $\xi$ is added to \code{OneLoopPotential} class, which can be set and get using, e.g.,
\begin{lstlisting}[language=c++]
double xi = model.get_xi();
model.set_xi(1); // set |\color{comment}$\xi = 1$| gauge
\end{lstlisting}
where \code{model} represents an instance of the  \code{OneLoopPotential} class. 
The Faddeev-Popov ghost contributions with non-zero $\xi$ are added using the third line of \cref{eq:MSOneLoopZeroTemp} and \cref{eq:Faddeev-Popov} when \code{get\_ghost\_masses\_sq} is provided.
For $\xi=0$,  \cref{eq: scalar mass} is implemented so that the user is spared from supplying the expression for the scalar masses.  Examples using \msbar scheme can be found in:
\begin{itemize}
    \item \code{EffectivePotential/include/models/2D\_test\_model.hpp} and
     \code{example/run\_2D\_test\_model.cpp}
    \item  \code{EffectivePotential/include/models/xSM\_MSbar.hpp} and
     \code{example/xSM/run\_xSM\_MSbar.cpp}
    \item \code{example/THDMIISNMSSMBCsimple/\linebreak{}THDMIISNMSSMBCsimple.hpp} and
    \code{example/THDMIISNMSSMBCsimple/\linebreak{}run\_THDMIISNMSSMBCsimple.cpp}
\end{itemize}

\subsubsection{\texorpdfstring{\msbar}{MS-bar} and the Covariant Gauge}

The gauge-fixing Lagrangian has the following form in the general $R_{\xi}$ gauge:
\begin{equation}
    \mathcal{L}_{\text{gf}} = - (\mathcal{F}^a)^2/ (2 \xi),
\end{equation}
where the gauge condition is $\mathcal{F}^a = \partial_{\mu}A^{\mu, a} - \xi \phi_i (g T^a \overline{\phi}_i ) = 0$ \cite{Patel_2011}. The parameter $\overline{\phi}$ can be taken to be the classical field, $\phi_{\text{cl}}$. However, this consequently implies that evaluating $V(\phi_{\text{cl}})$ for different choices of $\phi_{\text{cl}}$ is akin to changing the gauge --- making it impossible to perform calculations in a consistent gauge. This inconsistency has been raised before \cite{PhysRevD.46.2628, LAINE1994173}, and leads one to consider an alternative approach. One such choice is to employ the covariant gauge (also called the Fermi gauge)~\cite{LAINE1994173, PhysRevLett.113.241801, PhysRevD.91.016009}. In this case, the gauge-fixing Lagrangian has the following form \cite{Papaefstathiou:2020iag}:
\begin{equation}
    \mathcal{L}_{\text{gf}} = - \frac{1}{2 \xi_W} (\partial^{\mu} W_{\mu}^a )^2 - \frac{1}{2 \xi_B} (\partial^{\mu} B_{\mu} )^2.
\end{equation}
This addresses the fixed-VEV issue; however, without the cancellation afforded by the $\xi \partial_{\mu}A^{\mu} \phi$ cross term in the expansion of $(\mathcal{F}^a)^2$, the covariant gauge introduces mixing between the gauge and Goldstone sectors.

In the \msbar scheme, the functional form of the effective potential, as given in \cref{eq:MSOneLoopZeroTemp}, remains unchanged. However, the mass eigenstates that appear in the one-loop corrections are modified~\cite{Papaefstathiou:2020iag}. The fermion, Higgs, and gauge boson eigenstates are unchanged from their $R_{\xi}$ counterparts, whilst the two gauge-fixing parameters, $\xi_W$ and $\xi_B$, appear implicitly in the masses of the mixed Goldstone-ghost scalar particles, which arise due to the additional Goldstone-gauge mixing terms. These new eigenstates have the form:
\begin{align}
    m_{1, \pm}^2 & = \frac{1}{2} \left( \chi \pm \Re \sqrt{\chi^2 - \Upsilon_W} \right), \\
    m_{2, \pm}^2 & = \frac{1}{2} \left( \chi \pm \Re \sqrt{\chi^2 - \Upsilon_Z} \right),
\end{align}
where the parameter $\chi$ is defined in terms of physical Higgs fields $\phi_h$ as:
\begin{equation}
    \chi = \frac{1}{\phi_h} \frac{\partial V_0}{\partial \phi_h}.
\end{equation}
The implicit dependence on $\xi_W$ and $\xi_B$ comes from the $\Upsilon$ terms, which are themselves proportional to the derivative of the tree-level potential. Hence, any $\xi$ dependence is removed when evaluated in the tree-level minimum, as expected from Nielsen's identities \cite{nielsen_1975, Fukuda:1975di}. Furthermore, for sufficiently large $\xi$, it is possible that the arguments of the above square roots can become negative. In these cases, taking the real component effectively vanquishes any $\xi$ dependence.

The implementation of model in covariant gauge is similar to the one in $R_\xi$ gauge, both using the \code{OneLoopPotential} class. 
It is adopted in 
\begin{itemize}
    \item \code{EffectivePotential/include/models/xSM\_MSbar.hpp}\\
     \code{example/xSM/run\_xSM\_MSbar.cpp}
\end{itemize}
when the flag \code{use\_covariant\_gauge} is set to \code{true}.

\subsubsection{On-Shell Like Scheme}

As an alternative to the \msbar scheme, On-Shell (OS) like schemes for the effective potential \cite{Anderson_Hall_1992} are often used in phase transition calculations.  In OS-like schemes renormalisation conditions are chosen to fix the parameters so that the physical masses and mixings are given by the tree-level values, computed from the tree-level potential. Such schemes have the advantage that one may fix parameters from data without an iteration that is required in \msbar scheme and can be simple to implement if the counter terms are already known.\footnote{On the other hand it has the disadvantage that parameters will then be defined in a different renormalisation scheme to many automated calculations or observables where the \msbar scheme is commonly used, making it harder to connect with other observables. Therefore we allow both OS-like and \msbar schemes in \PT}

The counter terms are fixed from the renormalisation conditions to enforce vanishing loop corrections to tadpoles and physical masses in the effective potential approximation, i.e.\ the renormalisation conditions impose vanishing single and double derivatives, 
\begin{equation}\label{eq:os_condition}
    \left.\partial_{\phi_i} \overline{V_1} \right|_{\phi_i = v_i}= 0, \quad \left.\partial^2_{\phi_i\phi_j} \overline{V_1} \right|_{\phi_k = v_k} = 0.
\end{equation}
Here the overlined $\overline{V_1} = V_\text{CW} + V_\text{CT}$  denotes the zero temperature one-loop corrections which are the sum of the Coleman-Weinberg potential $V_\text{CW}$ and the counter term potential $V_\text{CT}$.  The latter should be of the form,
\begin{align}
    V_\text{CT} = \sum_i (\partial_{p_i}V_0) \, \delta p_i + \sum_k \delta T_k(\phi_k + v_k).
\end{align}
where $p_i$ are the parameters in the potential with corresponding counter terms $\delta p_i$ and $\delta T_k$ are the counter terms for the tadpoles.  

OS-like schemes of this from are employed in the \code{BSMPT} code \cite{Basler_2019,Basler:2020nrq,basler2024bsmptv3toolphase}, and already since first release of \pt it has been possible to use potentials from \code{BSMPT} in \pt.  However with \code{PhaseTracer2} we now include native \pt potentials in the OS-like scheme. 

In the Landau gauge if all masses have a simple field dependence of the form,   
\begin{equation}
    m^2(\phi) = m^2 + g \phi^2
    \label{eq:MassCons}
\end{equation}
then OS-like renormalisation conditions (\cref{eq:os_condition}) can lead to the following one-loop potential at zero temperature~\cite{Anderson_Hall_1992}:
\begin{equation}
    \begin{split}
    \overline{V_1}(\phi) = & \frac{1}{4 (4 \pi)^2} \sum_i n_i (-1)^{2 s_i} \bigg [ 2 m_i^2(\phi) m_i^2(v) \\ & + m_i^4(\phi) \left(\log \left( \frac{m_i^2(\phi)}{m_i^2(v)} \right) - \frac{3}{2}  \right)  \bigg ],
    \end{split}
    \label{eq:cwOSlike}
\end{equation}
where the summation again runs over all species in the mass spectrum.  This form for the potential has been commonly employed in extensions of the standard model with higher dimensional operators or a new scalar singlet.  Therefore in \code{PhaseTracer2}, we provide an example of this for the scalar singlet model\footnote{Note in general the scalar singlet model may not satisfy the requirement that all masses are written in the form of \cref{eq:MassCons} due to mixing between the scalars.  However it is satisfied for vacua where the VEVs are $(0, v_s(T) )$ and $(v_h(T), 0)$, and phase transitions to or between these vacua are often of interest.},
\begin{itemize}
    \item \code{EffectivePotential/include/models/xSM\_OSlike.hpp}\\
    \code{example/xSM/run\_xSM\_OSlike.cpp}
\end{itemize}
where the \code{V1} function of the \code{OneLoopPotential} class is overriden to use \cref{eq:cwOSlike}. 

For models that do not have such a simple dependence for the field dependent masses one can instead construct potentials where the counter terms are added to achieve the OS conditions \cref{eq:os_condition}. This can be done using the \code{counter\_term} function in the \code{OneLoopPotential} class which has been added as part of the \code{PhaseTracer2} extension. An example of this can be found in 
\begin{itemize}
    \item \code{example/THDM/THDM.hpp}\\
    \code{example/THDM/run\_THDM.cpp}
\end{itemize}
for the two Higgs doublet model~(THDM). 

\subsection{High Temperature Resummations}
\subsubsection{Daisy Resummation}

Finite temperature field theory is known to suffer from an IR problem \cite{weinberg1974gauge, dolan_jackiw_1974}, in which the perturbative expansion is expected to break down at the critical temperature. For example, in an interacting scalar theory the loop expansion of the effective potential contains daisy diagrams which dominate at each loop order. These introduce temperature enhancement to the scalar self-coupling $\lambda \to \lambda T^2/m^2$, which then diverges in the regime $ T \gg m$. This is alleviated by resumming these diagrams and introducing a temperature-dependent mass counter-term, known as daisy resummation.

There exist two prescriptions for inclusion of these resummed corrections. Both methods of resummation are shown in Ref.~\cite{athron2022arbitrary}, and in \code{PhaseTracer2} they are implemented for the scalar singlet extension (see \cref{sec:ztwo_ssm}). To briefly summarise, there is:
\begin{itemize}
    \item The Parwani procedure \cite{parwani1992resummation}, in which the mass eigenstates appearing in, e.g., \cref{eq:MSOneLoopZeroTemp} and \cref{eq:oneloopFiniteTemp}, are replaced with their resummed counterparts: \begin{equation}
        m_i^2(\phi) \rightarrow m_i^2(\phi) + \Delta_i,
    \end{equation}
    where $\Delta_i$ is the thermal Debye mass, which can be obtained from the finite-temperature self-energy. This replacement is performed for all scalars and longitudinal gauge bosons.
    \item The Arnold-Espinosa procedure \cite{arnold1993effective}, in which the following daisy correction term is included in the potential:
    \begin{equation}
        \begin{split}
                V_{\text{daisy}}(\phi, T) & = - \frac{T}{12 \pi} \times \\ 
               &  \sum_i  \bigg [ (m_i^2(\phi, T))^{3/2} - (m_i^2(\phi))^{3/2}\bigg ],    
        \end{split}
    \end{equation}
    where the first of these masses, $m_i^2(\phi, T)$, includes Debye corrections.
\end{itemize}

In the original version of \PT only the Arnold-Espinosa approach to daisy resummation was available and was included in the one-loop potential whenever the user provided Debye masses.  In \code{PhaseTracer2} the Parwani method is now included and the user can choose which daisy resummation method (if any) to use by utilizing one of the following lines of code:
\begin{lstlisting}
model.set_daisy_method(EffectivePotential::DaisyMethod::None);
model.set_daisy_method(EffectivePotential::DaisyMethod::Parwani);
model.set_daisy_method(EffectivePotential::DaisyMethod::ArnoldEspinosa);
\end{lstlisting}
where \code{model} represents an instance of the  \code{OneLoopPotential} class.  By default \code{DaisyMethod} is set to \code{ArnoldEspinosa}, though as before this will only change the potential to include high temperature resummation when the Debye masses are included.  Daisy resummation can can be used in any of the examples distributed with \PT where thermal masses are provided.

\subsubsection{3D Effective Field Theory} \label{sec:3dEFT}

Although Daisy resummation can mitigate infrared divergences to some extent, the 4D effective potential still suffers from strong gauge dependence, significant sensitivity to the renormalization scale, and slow convergence. These issues highlight the considerable limitations of 4D perturbative theory. Since the thermal plasma naturally exhibits a hierarchy of mass scales at high temperatures, perturbation theory can describe this by integrating out the contributions from the heavy thermal scale to the parameters of an effective theory at lower scales. This insight suggests that effective descriptions, particularly 3DEFT, can be used to obtain a more reliable effective potential. Significant progress has recently been made in this direction. Ref.~\cite{Kainulainen:2019kyp} demonstrated the shortcomings of conventional perturbative approaches based on the resummed effective potential in the two-Higgs doublet model (2HDM) and showed that 3DEFT results are largely insensitive to the renormalization scale. Subsequently, Ref.~\cite{Croon:2020cgk} explored the impact of renormalization scale dependence on gravitational wave predictions and further emphasized the advantages of 3DEFT, including its independence from gauge choices. Building on those foundations, a minimal 3DEFT approach that reconciles both gauge invariance and thermal resummation was proposed and successfully applied to the complex singlet extension of the Standard Model~\cite{Schicho:2022wty}. Furthermore, by applying 3DEFT separately to two-step phase transitions, the resulting thermodynamic properties were found to agree remarkably well with non-perturbative lattice simulations~\cite{Gould:2023ovu}.

Since phase transitions are inherently non-perturbative processes, the extent to which 3DEFT can faithfully reproduce lattice results remains an open question. To verify it, Ref.~\cite{Niemi:2021qvp} demonstrated that in the real-singlet extension of the Standard Model, the results of 1-loop and 2-loop 3DEFT calculations exhibit at least a 20\% discrepancy in critical temperature and latent heat, with differences reaching up to 100\% for certain two-step phase transition parameter points. This highlights the critical importance of including 2-loop effects when studying the thermodynamic properties of the system. Recently, Ref.~\cite{Niemi:2024axp, Niemi:2024vzw} further validated the consistency between 2-loop 3DEFT results and non-perturbative lattice simulations. Meanwhile, Ref.~\cite{Ekstedt:2022zro} introduced the `$x$-expansion' technique, which further reduced the intrinsic uncertainties of 3DEFT and showed that the infamous Linde problem is not as severe in these calculations as previously thought. The results for 3-loop 3DEFT have also been recently obtained, pushing the accuracy of this approach to a new level~\cite{Ekstedt:2024etx}.


Below, we will briefly introduce how the dimensionally-reduced effective field theory is constructed. The central tenet is that a finite-temperature $d+1$ dimensional relativistic field theory is equivalent to a $d$ dimensional Euclidean field theory with an additional compact dimension. Thus, a dimensional reduction can be performed on this compact dimension to yield an effective $d$ dimensional Euclidean field theory, which is matched to the parent $d+1$ dimensional theory. Equilibrium thermodynamics can then be derived from this simpler effective theory. For example, consider a real scalar theory given by the following action:
\begin{equation}
    i S = i \int d^{d+1} x \ \bigg (\frac{1}{2} (\partial_{\mu} \phi)(\partial^{\mu} \phi) - \frac{1}{2} m \phi^2 \bigg ),
    \label{eq:MinkowskiAct}
\end{equation}
where $x = (x^0, \mathbf{x})$. Upon compactification of the $x^0$ direction, $\mathbb{R}^{d + 1} \rightarrow \mathbb{R}^d \times S^1$, we Fourier decompose $\phi$ as:
\begin{equation}
    \phi( x^0, \mathbf{x} ) = \sum_n \phi_n(\mathbf{x}) e^{ 2 n \pi i \tau/ \beta }.
\end{equation}
where $\tau = i x^0$ and $\beta=1/T$. With this, \cref{eq:MinkowskiAct} becomes:
\begin{equation}
    \begin{split}
    i S = T \sum_n \int & d^d \mathbf{x} \ \bigg (\frac{1}{2} (\nabla_d \phi_n )^2 \\ 
    & + \frac{1}{2} [ m^2 + (2 \pi n T)^2  ]\phi_n^2 \bigg ),
    \end{split}
\end{equation}
At this point, the dimensional reduction is performed via summation over $n$.

Prior to performing the dimensional reduction, one introduces a hierarchy of scales to the thermal masses of all modes present in the parent theory. At the super-heavy scale come modes with thermal masses proportional to $\pi T$, which include all fermionic modes and non-zero bosonic modes (that is, modes for which $n \ne 0$ above). At the heavy scale are thermal masses proportional to $g T$, including all temporal gauge fields. Finally, the light scale contains modes with thermal masses proportional to $g^2 T$, including all zero scalar fields.\footnote{Note the labels hard, soft, and ultrasoft are often used (for example by \code{DRalgo}) in place of super-heavy, heavy, and light respectively.} Spatial gauge fields are always light.

The dimensional reduction is performed by matching, taking place in two steps. Given the full $d+1$-dimensional Lagrangian, one integrates out the hard modes by writing the most general $d$-dimensional super-renormalisable Lagrangian containing only the soft and ultrasoft modes, in which all Lagrangian parameters are unknown functions of temperature. The 2-, 3-, and 4-point 1PI Green's functions, $G$, are then calculated in both theories, and the unknown effective parameters are set to match the results of both theories. Once the matching is performed, the effective potential can then be derived within the reduced theory.

Restricting to the conventional 4D case, we summarise this process below:
\begin{equation}
    \mathcal{L}_4 \rightarrow \mathcal{L}_3 \rightarrow V_{3, \text{eff}}.
\end{equation}
\code{PhaseTracer2} includes examples for both the Abelian Higgs and xSM models, both of which we obtained using the code \code{DRalgo} \cite{Ekstedt_2023}. We employ the minimal scheme laid out in Ref.~\cite{Schicho_2022}, where in the dimensional reduction is performed to NLO, whilst the effective potential is calculated to LO. That is, during the matching procedure masses are calculated to two-loop order, whilst couplings are determined to one-loop order. The effective potential is then calculated to one-loop: 
\begin{equation}
    V_{3, \text{eff}} (\phi_3, T) = V_{3, 0} + V_{3, 1},
\end{equation}
where $V_{3, 0}$ is the tree-level potential (equivalent to the 4D case but with the replacements from $\mathcal{L}_4 \rightarrow \mathcal{L}_3$) and in three dimensions the one loop correction is UV finite and given by \cite{Lewicki:2024xan}:
\begin{equation}
    V_{3, 1} = -\frac{1}{12 \pi} \sum_{i} n_i [ m_i^2(\phi_3)]^{3/2}.
    \label{eq:3d_1_loop}
\end{equation}
We calculate the potential in the ultrasoft regime, such that all fermionic degrees of freedom, together with temporal gauge boson contributions, have been integrated out. As such, the sum in \cref{eq:3d_1_loop} includes both scalar masses and longitudinal gauge bosons. Additionally, $n_i = 1$ for scalar contributions and $n_i = (d - 1) \rightarrow 2$ for longitudinal gauge bosons.

In a bare-bones approach, with $V_{3, \text{eff}} = V_{3, 0}$ and LO matching, dimensional reduction is identical to the high temperature expansion. To elucidate, consider a real scalar $\phi^4$ theory with mass $m$ and quartic coupling $\lambda$. Expanding in powers of $\lambda$, to leading order the three dimensional mass has the following one-loop expression:
\begin{equation}
    m_3^2 = m_4^2 + c_\phi T^2,
    \label{Eq:msq_match}
\end{equation}
where all terms are leading order in the couplings, i.e.\ of order $\lambda \sim g^2$, and the subscripts indicate quantities from both the three dimensional and parent four dimensional theories.  The last term in Eq.\ \ref{Eq:msq_match} is obtained from taking only the leading order contribution of the one-loop finite temperature potential, meaning that $c_\phi$ must the Debye coefficient. 
Likewise, at leading order the equivalent $\mathcal{O}(\lambda)$ expression for the quartic coupling is\footnote{For the expressions of both $m_3^2$ and $\lambda_3$ provided here, compare to the equivalent $\mathcal{O}(g^2)$ expressions given in eqs.~(1.3) and~(1.4) of Ref.~\cite{Ekstedt_2023}.}:
\begin{equation}
    \lambda_3 = \lambda_4 T.
\end{equation}
The quartic coupling in the 3D theory gains a positive mass dimension, reflecting the new dimension of the scalar field, $[ \phi_3 ] = 1/2$. Lastly, on dimensional grounds we can make the relations: $V_4 = T V_3$ and $\phi_4 = \sqrt{T} \phi_3$. Hence:
\begin{equation}
    \begin{split}
        V_4 & = T V_3 \\
        & = T \left [ m_3^2 \phi_3^2 + \lambda_3 \phi^4_3 \right ] \\
        & = T \left [ (m_4^2 + c_{\phi} T^2 ) \phi_4^2 T^{-1} + (\lambda_4 T) \phi^4_4 T^{-2} \right ] \\
        & = (m_4^2 + c_{\phi} T^2 ) \phi_4^2 + \lambda_4 \phi^4_4, \\
    \end{split}
\end{equation}
This is the same result obtained in the high temperature expansion, which we discuss in \cref{sec:HighTemperature}. 

\subsubsection{DRalgo} 
\label{sec:DRalgoExp}

As mentioned above, \code{PhaseTracer2} contains the 3D effective field theory (3DEFT) for both the Abelian Higgs and xSM models, which can be found in \cref{sec:DRalgo_ah} and \cref{sec:dim_reduced_ssm} respectively. These examples were built with \code{DRalgo}. \code{DRalgo} is a Mathematica package designed to automate the process of construction a dimensionally reduced theory. Starting from a basic Lagrangian input, it can be used to obtain the 4D beta functions, all 3D thermal masses and couplings at both the soft and ultrasoft scales, and up to the two-loop effective potential in the 3D theory. This process is cumbersome to do by hand, and given the recent advancements in utilising the 3DEFT for cosmological phase transition studies, having such an interface between \code{DRalgo} and \code{PhaseTracer2} will open the door for further studies into not only extended scalar sectors of the SM, but also new gauge groups and beyond. We will refer the reader to the examples section for details on running each  example program, which we will not detail here. Instead, in this short section, we outline how we utilise the output from \code{DRalgo} to construct our code.

Firstly, the relevant model classes are created by overriding the virtual \code{EffectivePotential::Potential} class. This requires additional functionality in the form of the \code{solveBetas} and \code{get\_3d\_parameters} functions. The method \code{solveBetas} is used to numerically solve the one-loop beta functions in the parent 4D theory, which facilitates the running of 4D input parameters to the scale of the dimensional reduction,  $\mu_{\text{4D}} = \pi T$. In order to use \code{solveBetas}, the user must provide the relevant couplings and beta functions using the \code{Betas} function, which has the syntax:
\begin{lstlisting}[language=c++]
void Betas(const std::vector<double>& x, std::vector<double>& dxdt, const double t) {
  double gsq   = x[0];  // |\color{comment}$g^2$|
  double muhsq = x[1];  // |\color{comment}$\mu_h^2$|
  double lam   = x[2];  // |\color{comment}$\lambda$|
  dxdt[0] = 1/t * ...;  // |\color{comment}$\beta_{g^2}$|
  dxdt[1] = 1/t * ...;  // |\color{comment}$\beta_{\mu_h^2}$|
  dxdt[2] = 1/t * ...;  // |\color{comment}$\beta_{\lambda}$|
}
\end{lstlisting}
where \code{x} contains the 4D parameters, and \code{dxdt} provides the corresponding beta functions.\footnote{Note they are passed by non-constant reference for interoperability with \code{boost} ODE routines.} We have used the relevant parameters from the Abelian Higgs model, and for brevity indicated the definitions of the beta functions by \code{...}. Both the expressions for the beta functions, and the expressions for the 3D parameters below, are implemented by hand using a manual conversion of the Mathematica language. The \code{solveBetas} solutions are stored in the member variable \code{RGEs}.

Once we have run the 4D parameters from the input scale $Q_0$ to the relevant dimensional reduction scale $\mu_{\text{4D}} = \pi T$, we perform said dimensional reduction using \code{get\_3d\_parameters}. This takes the 4D parameters and matches them to their 3D counterparts, which are then returned by the function. That is, \code{get\_3d\_parameters} might have the following form:
\begin{lstlisting}[language=c++]
std::vector<double> get_3d_parameters(double T) const {
  double mu4d = M_PI * T;  // |\color{comment}$\mu_{\textrm{4D}}$|
  double gsq4d = alglib::spline1dcalc(RGEs[0], mu4d); // |\color{comment}$g^2_{\textrm{4D}}$|
  double gsq3d = gsq4d * T;  // |\color{comment}$g^2_{\textrm{3D}}$|
  return {gsq3d};
}
\end{lstlisting}
We first define $\mu_{\text{4D}}$, evaluate any relevant 4D parameters at $\mu_{\text{4D}}$ (e.g.~$g^2_{\text{4D}}$), then use these parameters to evaluate their 3D counterparts (e.g.~$g^2_{\text{3D}}$) which are then returned. Note that for the particular case of gauge couplings $g$, we use $g^2$ instead of $g$ in accordance with the conventions used by \code{DRalgo}. Not shown above is that the 3D parameters can also depend on the additional parameters $\mu_{\text{3D}}$ and the ultrasoft $\mu_{\text{3D-US}}$. These arise from RG running within the 3D theory, required to resum logarithms. In our code, we choose to set these as:
\begin{equation}
    \mu_{\text{3D}} = g T, \quad \mu_{\text{3D-US}} = g^2 T,
\end{equation}
which suppresses logarithms of the form 
\begin{equation}
    \ln \left( \frac{\mu_{\text{3D}}}{m_G} \right)
\end{equation}
where $m_G$ are the various gauge boson Debye masses, and these logarithms arise in the 2-loop expressions for our 3D masses. As couplings are only evaluated at most 1-loop order, we need not worry about the appearance of logarithms in those parameters.

\begin{figure}
\centering
\begin{tikzpicture}[node distance=0.8cm,font=\sf]
\node (a) [box] {Define 4D parameters $\{m_{\text{4D}}^2, g_{\text{4D}}^2, \lambda_{\text{4D}} \}$ at initial scale $Q_0$};
\node (b) [box, below=of a] {RG-evolve the 4D parameters down to $\mu_{\text{4D}} = \pi T$ using \code{solveBetas}};
\node (c) [box, below=of b] {Match RG-evolved 4D parameters to their 3D counterparts $\{m_{\text{3D}}^2, g_{\text{3D}}^2, \lambda_{\text{3D}} \}$ using \code{get\_3d\_parameters}};
\node (d) [box, below=of c] {Use 3D expressions in the effective potential};
\draw [arrow] (a) -- (b);
\draw [arrow] (b) -- (c);
\draw [arrow] (c) -- (d);
\end{tikzpicture}
\medskip
\caption{Outline of 3DEFT methodology.}
\label{fig:3DEFT}
\end{figure}
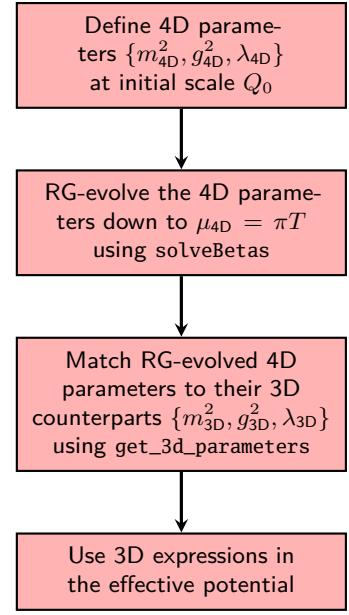

Lastly, we utilise \code{get\_3d\_parameters} when defining our potential. Much of the new functionality of \code{PhaseTracer2} pertaining to the bounce action and gravitational waves is designed with the 4D potential in mind. As such, when we define the 3D potential we must take care to make the required conversions:
\begin{equation}
    V_4 = T V_3, \quad \phi_4 = \phi_3 \sqrt{T}.
\end{equation}
We summarise the methodology we employ in \cref{fig:3DEFT}.

\subsection{Gauge Invariant Potentials}
\subsubsection{High Temperature Expansion}
\label{sec:HighTemperature}

The thermal functions that arise in $V_T(\phi, T)$ cannot be expressed in terms of elementary functions. However, prescriptions for including these terms exist, such as numerical integration or tabulating values.  The latter option is utilised by the \code{EffectivePotential} library. Alternatively, these functions admit both low-temperature (LT) $z \equiv m/T \gg 1$ and high-temperature (HT) $z \ll 1$ expansions. To leading-order, the HT expansion has the form \cite{dolan_jackiw_1974}:
\begin{align}
    \frac{T^4}{2 \pi^2} J_B(z^2) &  \simeq - \frac{\pi^2 T^4}{ 90 } + \frac{m^2 T^2}{24} \\
    \frac{T^4}{2 \pi^2} J_F(z^2) & \simeq \frac{7 \pi^2 T^4}{720} - \frac{m^2 T^2}{48}.
\end{align}
Thus, by omitting the zero-temperature CW corrections, these expansions allow for a simple high-temperature approximation to the potential, which is equivalent to the following replacement of tree-level masses:
\begin{equation} \label{eq:DebyeMass}
    m_i^2 \to m_i^2 + c_i T^2,
\end{equation}
where the Debye coefficients $c_i$ depend on the high temperature expansion. E.g., for a $\phi^4$ theory of a single real scalar singlet, $c_i = \lambda \pi^2/12$, whilst for a Standard Model extension $c_i$ gains additional corrections due to both gauge boson and Yukawa contributions \cite{Vaskonen:2016yiu}. The simplicity of this prescription means that the potential in this scheme is an, at most, quartic polynomial in the various scalar fields. Consequently, analytic expressions for the critical temperature and transition strength can often be obtained.

As a result of the expansion of the thermal functions, the gauge fixing parameter $\xi$ no longer explicitly appears in the potential. Note here that nothing has been done to achieve genuine gauge invariance, and this is simply an artefact of the approximation employed. Additionally, in discarding the CW contributions to the potential we no longer have explicit dependence on the regularisation scale $Q$. Thus at first glance, this expansion has liberated any explicit $\xi$ or $Q$ dependence. However, the lack of an explicit $Q$ in the potential may worsen scale dependence, as explicit and implicit scale dependence can be expected to partially cancel. These factors, together with this limiting case only applying in the $m/T \ll 1$ regime, should be treated with care and caution when applying the high temperature expansion.

Since the high-temperature expansions are relatively simple, they can be directly implemented utilizing the \code{Potential} class, as illustrated in
\begin{itemize}
    \item \code{EffectivePotential/include/models/Z2\_scalar\_singlet\_model.hpp}\\
    \code{example/scan\_Z2\_scalar\_singlet\_model.cpp}
    \item \code{EffectivePotential/include/models/xSM\_HT.hpp}\\
    \code{example/xSM/run\_xSM\_HT.cpp}
\end{itemize}
Additionally, in \code{PhaseTracer2}, the \code{OneLoopPotential} class automatically computes the high-temperature expansions when the thermal parts of \cref{eq:DebyeMass} are supplied by the \code{get\_scalar\_thermal\_sq} function, such as shown in 
\begin{itemize}
    \item \code{EffectivePotential/include/models/xSM\_base.hpp}
\end{itemize}
This computation will be integral to the $\hbar$ expansion method, as discussed in the next \cref{sec:hbar}. 

\subsubsection{The \texorpdfstring{$\hbar$}{h-bar} Expansion }
\label{sec:hbar}

The exact effective potential is gauge invariant when evaluated at a minimum, as shown by Nielsen's identities~\cite{nielsen_1975, Fukuda:1975di}. This ensures that some thermal parameters, such as the critical temperature, are gauge invariant.\footnote{The critical temperature is inherently defined with respect to the minimum of the potential, and is thus gauge invariant.} However, gauge dependence can arise in these calculations due to incorrectly terminating perturbative expansions. For example, below it is shown that the effective potential truncated at one-loop is gauge invariant when evaluated at the tree-level minimum. As such, gauge dependence arises if one instead evaluates at the one-loop minimum. The $\hbar$-expansion \cite{Laine:1994zq,Patel_2011} provides a means of determining the critical temperature and transition strength using perturbation theory whilst maintaining gauge invariance.

In this approach one writes both the potential and minimum as an expansion in the parameter $\hbar$:
\begin{align}
    V(\phi, T) & = V_0(\phi) + \hbar V_1(\phi, T, \xi) + \hbar^2 V_2(\phi, T, \xi) + \dots \\
    \phi_{\text{min}} & = \phi_0 + \hbar \phi_1(T, \xi) + \hbar^2 \phi_2(T, \xi) + \dots 
\end{align}
These are substituted into the minimisation condition
\begin{equation}
    \frac{\partial V}{\partial \phi} \bigg |_{\phi_{\text{min}}} = 0.
\end{equation}
The terms in this expression can then be arranged in powers of $\hbar$, and solving for $\phi$ order-by-order gives (up to $\mathcal{O}(\hbar)$):
\begin{align}
    \mathcal{O}(\hbar^0):& \quad\quad \quad \ \phi_0  = \frac{\partial V_0}{\partial \phi_0} \bigg |_{\phi_0} \\
    \mathcal{O}(\hbar^1):&\quad \phi_1(T, \xi) = - \bigg( \frac{\partial^2 V_0}{\partial \phi^2} \bigg)^{-1}_{\phi_0} \frac{\partial V_1}{\partial \phi_0} \bigg |_{\phi_0}.
\end{align}
Substituting these into the potential, one obtains the following expression for the minimum at each order of $\hbar$:
\begin{equation}
    \begin{split}
        V(\phi_{\text{min}}, T) & = V_0(\phi_0) + \hbar V_1(\phi_0, T, \xi) \\
        & + \hbar^2 \left[ V_2(\phi_0, T, \xi ) - \frac{1}{2} \phi_1^2(T, \xi) \left.\frac{\partial^2 V_0}{\partial \phi^2} \right|_{\phi_0 } \right].
    \end{split}
    \label{eq:hbarMinimum}
\end{equation}
Thus, the set of points $\phi_0$ that extremize the tree-level potential allow \cref{eq:hbarMinimum} to define a family of one-parameter $T$-curves showing how each phase evolves.\footnote{Note Nielsen's identity only requires $\partial V/\partial \phi = 0$, such that these points may be maxima or saddle points.} Any intersection point of these curves defines a potential critical point.

Hence, to obtain the critical temperature one first solves for tree-level minima $\phi_0$ (again saddle points and maxima can also be utilised, and indeed the one-loop vacuum structure may not resemble the tree-level case. In such instances we resort to saddle points and maxima). Any tree-level Lagrangian parameters are set using tree-level tadpoles and tree-level constraints on masses, ensuring gauge dependence does not arise unknowingly. Field-dependent mass eigenstates in the CW potential cannot be modified, nor can 
daisy resummation in the 4D theory be employed, to ensure Nielsen's identity remains valid. 

Once $\phi_0$ is obtained, \cref{eq:hbarMinimum} can be used at $\mathcal{O}(\hbar)$ via the one-loop potential without resummation to determine any temperatures with degenerate phases. Provided this degenerate point corresponds to a transition from one local minimum to another, this temperature corresponds to a critical temperature $T_c$. 

Using the  \code{HbarExpansion} class, the $\hbar$ expansion method can be used for any potential written in the $\msbar$ scheme with the \code{get\_scalar\_thermal\_sq} function provided~(See \cref{sec:HighTemperature}). The \code{HbarExpansion} class extends the \code{PhaseFinder} class.  To run \PT using the $\hbar$ expansion, an object of the \code{HbarExpansion} class is constructed from the potential in the same way a \code{PhaseFinder} object is constructed and then acted on in a similar way.  An example of this is given for the $Z_2$ symmetric scalar singlet extension: 
\begin{itemize}
    \item \code{EffectivePotential/include/models/xSM\_MSbar.hpp}\\
    \code{example/xSM/run\_xSM\_PRM.cpp}
\end{itemize}

\section{Bounce action and nucleation rate}\label{sec:TransProp}

In the first release of \code{PhaseTracer} the code could only be used to identify the possible phase transitions.\footnote{Details about how this can be done using the \code{PhaseTracer} code can be found in our previous paper~\cite{Athron:2020sbe}.} To determine which phase transitions complete one needs to compute and use the nucleation rate, i.e.\ the rate at which bubbles of the new phase form. The nucleation rate can be estimated by computing the bounce action and in this section we first briefly review the bounce action and how this can be used to estimate the nucleation rate.  We then describe how the bounce action can now be obtained within \code{PhaseTracer} where we include both native calculations (using a 1D shooting algorithim and path deformation for multiple fields) and an interface to the public bounce solver \code{BubbleProfiler}. Finally we give several examples of how the bounce action can be used including simple estimates of the nucleation rate.     

\subsection{Bounce action}

A first-order phase transition proceeds via the nucleation of true vacuum bubbles within the false vacuum plasma. The decay rate of a metastable phase is a well-established problem in many areas of physics, and evaluation of this decay rate is achievable through numerous equivalent formalisms. In the context of cosmological phase transitions, the most prevalent of these is Coleman and Callan's bounce formalism \cite{coleman1977fate, coleman1977fate2}, which involves finding an instanton solution to the Euclidean equations of motion that corresponds to a spherical bubble of the true vacuum.  

The decay rate is defined in terms of the generating functional $Z[J]$ by \cite{PhysRevD.109.045010}
\begin{equation} \label{eq: Z0}
    Z[0] = \langle \phi_{FV}| e^{- \mathcal{H} \mathcal{T}} | \phi_{FV} \rangle,
\end{equation}
for a (Euclidean) transition time $\mathcal{T}$. To derive this, from elementary quantum mechanics it is known the imaginary part of the potential energy corresponds to the decay rate of a particle:
\begin{equation}
    \Gamma = -2 \,\Im  E.
\end{equation}
To find $Z$, we write the propagator in Euclidean space as
\begin{equation}
	\bra{x_f}e^{-\mathcal{H}\mathcal{T}}\ket{x_i} = \sum_{n}e^{-E_n\mathcal{T}}\braket{x_f|n}\braket{n|x_i}
\end{equation}
where $\ket{x_i}$ and $\ket{x_f}$ are the initial and final states, respectively. For sufficiently large $\mathcal{T}$, this expansion is dominated by the ground state contribution, $ n = 0$. Thus,
\begin{equation}
	\bra{x_i}e^{-\mathcal{H}\mathcal{T}}\ket{x_i} \approx e^{-E_0\mathcal{T}}\braket{x_i|0}\braket{0|x_i} = C e^{-E_0\mathcal{T}},
\end{equation}
where all non-essential terms have been expressed by the real constant $C$. After we re-express the left-hand side of the equation as $Z$, $\Gamma$ can be written as 
\begin{equation}
    \Gamma = -2\, \Im E_0 = -2\Im  \left[-\frac{1}{\mathcal{T}}\ln Z +\frac{1}{\mathcal{T}} \ln C\right]
\end{equation}
Obviously, the part $\frac{1}{\mathcal{T}}\ln C $ is real and approaches to zero for large $\mathcal{T}$. Thus
\begin{equation} 
    Z[0] = \langle \phi_{FV}| e^{- \mathcal{H} \mathcal{T}} | \phi_{FV} \rangle,
\end{equation}
where we have changed $\ket{x_i}$ to $\ket{\phi_{FV}}$. The decay rate per unit volume is derived from this expression as:
\begin{equation}
    \frac{\Gamma}{V} = \lim_{\mathcal{T} \rightarrow \infty } \frac{2}{V \mathcal{T}} \, \Im \ln Z[0] .
\end{equation}
Utilising a steepest descent approximation Coleman and Callan showed this yields an expression of the form:
\begin{equation}
    \frac{\Gamma}{V} = A e^{-B} [ 1 + \mathcal{O}(\hbar) ],
    \label{Eq:Nuc_rate_zero_temp}
\end{equation}
or at finite temperature (courtesy of Linde) \cite{linde1981otherdecay, linde1983decay}:
\begin{equation}
    \frac{\Gamma}{V} = A(T) e^{-B(T)/T} [ 1 + \mathcal{O}(\hbar)
    ].
    \label{Eq:Nuc_rate_finite_temp}
\end{equation}
For decay via quantum fluctuations, we have $B = S_4$, where $S_d$ is the Euclidean action for $O(d)$-symmetric field configuration characterising a bubble connecting each phase. In general, this action is given by:
\begin{equation}
    S_d(\phi) = \Omega_{d - 1} \int_{0}^{\infty} \rho^{d - 1} \bigg ( \frac{1}{2} \dot{\phi}^2 + V(\phi) \bigg ),
\end{equation}
where $\Omega_n$ is the area of an $n$-sphere. The action is obtained by integrating over the bounce solution, $\phi_b(\rho)$, an instanton that interpolates between each phase, with $\rho = \sqrt{\tau^2 + |\mathbf{x}|^2 }$. $\phi_b$ is found by solving the equation of motion
\begin{equation}
    \ddot{\phi}(\rho) + \frac{d - 1}{\rho} \dot{\phi}(\rho) = V'(\phi),
\end{equation}
subject to the bounce boundary conditions:
    \begin{equation}
    \begin{split}
        \dot{\phi}(\rho) & = 0, \\
        \phi(\rho \rightarrow \infty) & = \phi_f, \\
        \dot{\phi}(\rho \rightarrow \infty) & = 0.
    \end{split}
\end{equation}
For decay via thermal fluctuations, we instead have $B(T) = S_3(T)$ and $\rho = |\mathbf{x}|$, otherwise the methodology is equivalent.

The dimensional part of the decay rate, $A$, characterises quantum fluctuations from the bounce solution. For zero-temperature tunnelling this is given by:
\begin{equation}
    A = \bigg ( \frac{S_4[\phi_b]}{2 \pi} \bigg )^2 \bigg ( \frac{ \det' [ - \nabla^2 + V''(\phi_b) ]}{ \det [ - \nabla^2 + V''(\phi_f) ] } \bigg )^{- \frac{1}{2}}.
\end{equation}
The primed functional determinant denotes omission of zero eigenvalues. Direct evaluation of these functional determinants is an onerous task, and they are often replaced with a relevant dimensional quantity \cite{linde1983decay}. At zero-temperature, one can use the radius of a critical bubble $R_0$, obtaining:
\begin{equation}
    A \sim R_0^{-4} \bigg ( \frac{S_4[\phi_b]}{2 \pi} \bigg )^2.
    \label{A_est_zero_temp}
\end{equation}
At finite temperature, $A(T)$ has equivalent expressions for both tunneling and thermal fluctuations over the barrier. Again utilising an approximation for the functional determinants, these become:
\begin{equation}
    A(T) \sim T^4 \bigg ( \frac{S_3[\phi_b(T) ]}{2 \pi T } \bigg)^{\frac{3}{2}}.
    \label{A_est_finite_temp}
\end{equation}
Recently, developments have been made toward accessing the functional determinant without needing to resort to these dimensional arguments. The first of its kind python package \code{BubbleDet} \cite{Ekstedt:2023sqc} allows for an evaluation of these functional determinants for both fluctuating scalar and gauge fields. The package is currently limited to the special case where the determinant can factorise into a product of individual determinants, but it nevertheless takes great strides toward removing the large theoretical uncertainty generated by the approximations above. Interfacing \code{PhaseTracer} with this code is an expected forthcoming feature. For now, \code{PhaseTracer2} just adopts the approximation that $A \sim T^4$.

Evaluation of the bounce action in \code{PhaseTracer2} is performed within the \code{ActionCalculator} class via the \code{get\_action} method. This evaluates the bounce action using either an improved version of the shooting and path deformation algorithms first used in \code{CosmoTransitions} \cite{wainwright2012cosmotransitions}, or by interfacing directly with \code{BubbleProfiler}. The user can specify this choice using one of the following flag setting:
\begin{lstlisting}
  ac.set_action_calculator(PhaseTracer::ActionMethod::BubbleProfiler);
  ac.set_action_calculator(PhaseTracer::ActionMethod::PathDeformation);
  ac.set_action_calculator(PhaseTracer::ActionMethod::All);
\end{lstlisting}
where \code{ac} represents an instance of the  \code{ActionCalculator} class. If \code{ActionMethod::All} is chosen, both the action calculators will be employed, and the smaller result will be utilized.

\subsection{One-dimensional Shooting Method}

In one dimension both \code{BubbleProfiler} and path deformation employ an overshoot/undershoot method proposed by Coleman~\cite{coleman1977fate}. An initial field configuration $\phi_0 \equiv \phi(\rho = 0)$ is considered and evolved in time, checking whether it overshoots --- indicating $\phi_0$ started too close to the true vacuum --- or undershoots --- indicating it started too close to the well between each maximum --- the false vacuum. 

To implement this in \code{PhaseTracer2}, we first extract the effective potential using an instance of the \code{OneDimPotentialForShooting} class, before passing this instance to the \code{Shooting} class. This class contains two methods that enable computation of the bounce action. First, \code{findProfile}, determines the field profile using the shooting method outlined above. Then, \code{calcAction} integrates this field profile over the bubble volume to determine the associated bounce action.

This method is used in 
\begin{itemize}
    \item \code{example/TestPathDeformation/run\_shooting\_1d.cpp}
    \item \code{example/TestBubbleProfiler/run\_BP\_scale.cpp}
\end{itemize}
for temperature-independent potentials without and with \code{BubbleProfiler}.

\subsection{Multi-dimensional Path Deformation Algorithm}

The path deformation method is based upon the notion of intrinsic coordinates from classical mechanics, originally utilised in \code{CosmoTransitions}~\cite{wainwright2012cosmotransitions}. We express the bounce equation in intrinsic coordinates by parameterising the path in terms of $x = x(\rho)$, such that $\phi \equiv \phi (x)$ and we define $x$ such that $|d\phi/dx| = 1$.\footnote{Here it is understood that $\phi \equiv (\phi_1, \dots, \phi_n)$.} In this way, $d \phi/ d x = \hat{e}_t(x)$ is a unit tangent vector to the path. Then, we obtain:
\begin{align}
    \dot{\phi} & = \dot{x} \hat{e}_t(x), \\
    \ddot{\phi} & = \ddot{x} \hat{e}_t(x) + \dot{x}^2 \bigg | \frac{d^2 \phi}{d x^2} \bigg | \hat{e}_n(x),
\end{align}
where $\hat{e}_n(x)$ is unit normal to the tangent vector to the path. In these coordinates, the bounce equation becomes:
\begin{equation}
    \ddot{x} \hat{e}_t(x) + \dot{x}^2 \bigg | \frac{d^2 \phi}{d x^2} \bigg | \hat{e}_n(x) + \frac{d-1}{\rho} \dot{x} \hat{e}_t(x) = \nabla V(\phi(x)),
\end{equation}
which splits into the following equations:
\begin{align}
    \ddot{x} + \frac{d-1}{\rho} \dot{x} & = (\nabla \cdot \hat{e}_t(x) ) V(\phi(x)) = \frac{d V(\phi(x))}{d x}, \\
    \dot{x}^2 \bigg | \frac{d^2 \phi}{d x^2} \bigg | & = (\nabla \cdot \hat{e}_n(x)) V(\phi(x)), \\
    0 & = (\nabla \cdot \hat{e}_b(x))) V(\phi(x)).
\end{align}
That is, the equation splits into three groups: one defining the motion along field trajectory, one defining the motion normal to this, and an additional set of $n-2$ equations for the binormal directions. The path deformation method proceeds as follows: an initial guess solution of the form $\phi(x)$ is proposed, typically taken to be a straight line in field space connecting the two phases. The first $x$ direction equation is then solved using the shooting method to derive $x(\rho)$. Then, given this path one computes the normal force, $F_n(x)$, required to ensure the normal equation is satisfied. If this vanishes, clearly we have obtained the correct path. If this is non-zero, we have obtained the incorrect path, and we then perturb the path in the normal direction and proceed iteratively. That is, we determine the new perturbed path, calculate the normal force, and check to see if it vanishes. Once we obtain a vanishing normal force we have found the correct bounce solution. Provided we have torsion and the tangent, normal, and binormal directions will all change each iteration. Hence, the binormal equations are satisfied at each step.

The path deformation method is implemented in \code{PhaseTracter2} by creating an instance of the \code{PathDeformation} class. Here the bounce action is evaluated using the method \code{full\_tunneling}. This method evaluates both the field profile and the bounce action using the path deformation method outlined above. The field profile and action --- amongst other things --- are then stored within the \code{struct} \code{full\_tunneling}. The bounce action can then be accessed using the method \code{get\_action}.

There are two temperature-independent 2D potentials in
\begin{itemize}\sloppy
    \item \code{example/TestPathDeformation/run\_pathdeformation\_2d.cpp}
    \item \code{example/TestPathDeformation/run\_pathdeformation\_BP\_2d.cpp}
\end{itemize}
for testing the path deformation method. 

\subsection{BubbleProfiler}

In order to maintain robustness, \code{PhaseTracer2} features an interface with \code{BubbleProfiler}~\cite{Athron_2019}. This provides an alternative means of calculating the bounce action, and importantly, a means of doing so via a different algorithm. Here we briefly summarise the method employed by \code{BubbleProfiler}, and refer the reader to \cite{Athron_2019} for more detail. 

To solve the one dimensional problem, \code{BubbleProfiler} again utilises the shooting method where an initial guess of the field and velocity are evolved using a controlled Runge-Kutta like method. Unlike \code{PhaseTracer2} and \code{CosmoTransitions}, however, the multi-field problem is addressed using a  perturbative method. First, an initial guess is made, typically assumed to be a straight line in field space. The profile along this guess is then either obtained from solving the one-dimensional shooting method along this line, or by using an analytical ansatz motivated by a fourth-order polynomial approximation of the potential. 

From this initial ansatz, corrections are then calculated perturbatively. That is, for each field direction $\phi_i$, the $j$-th correction is found as $\phi_i^{(j)} = \phi_i^{(j-1)} + \epsilon_i^{(j-1)}$. The perturbations $\epsilon_i$ are found by linearising the equations of motion. This reduces the problem from solving $N$ non-linear coupled PDEs, to solving $N$ linear coupled PDEs --- a categorically simpler problem to treat numerically. Notably this method has the shortcoming that convergence is dependent on an adequately correct initial guess. Provided such a guess is made, the algorithm is terminated once changes in both the initial starting position and the Euclidean action each iteration fall below a certain threshold.

The BVP for the corrections $\epsilon_i$ is solved using a generalisation of the shooting method. As the problem is now linear, such a generalisation becomes both efficient and tractable. Indeed, the algorithm converges within a single step. 

Testing has shown both the interface with \code{BubbleProfiler}, and the path deformation method we provide have good agreement. For concrete evidence, see the examples given in \cref{sec:ExamplePolynomial,sec:Example2D}. Additionally, see
\begin{itemize}
    \item \code{example/TestBubbleProfiler/run\_BP\_2d.cpp}
\end{itemize}
for a simple example for calling \code{BubbleProfiler}

\subsection{Dealing with discrete symmetries}\label{sec:discrete symmetries}
Before we look at applications using the bounce action discussed in the previous subsections,  we first discuss an important subtle point for transitions between phases related by discrete symmetries, and clarify how we handle this in our action calculation.  An important distinction between \code{PhaseTracer} and \code{CosmoTransitions} is the treatment of discrete symmetries. While \code{CosmoTransitions} utilizes a user-defined function to carve out the symmetric field space, \code{PhaseTracer} thoroughly examines all conceivable transitions between symmetric phases, as elaborated in the first manual~\cite{Athron:2020sbe}.  This will also result in differences in the action calculation.

\begin{figure}
    \centering
    \includegraphics[width=\linewidth]{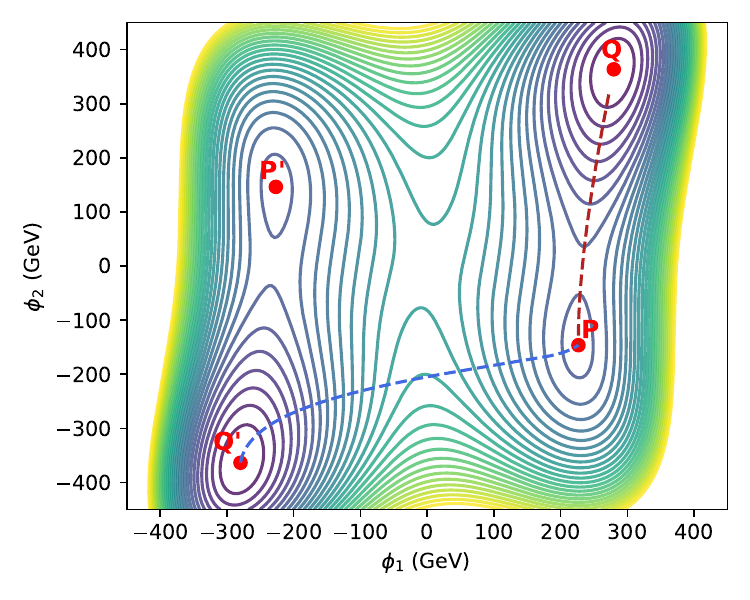}\\
    \includegraphics[width=\linewidth]{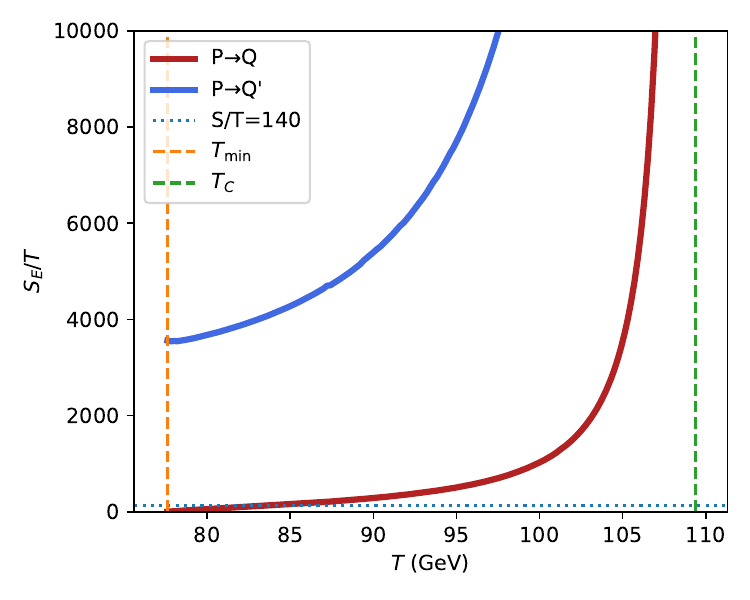}
    \caption{Top: Potential of the two-dimensional test model at $T=95\gev$. $P$ and $Q$ denote the minimums of the potential, while $P'$ and $Q'$ represent their symmetric partner. The dashed red and blue lines indicate the tunnelling paths from $P\to Q$ and $P\to Q'$, respectively.  Bottom: Action as a function of temperature between $P$ and $Q$ (red) and between $P$ and $Q'$ (blue). }
    \label{fig:discrete symmetries}
\end{figure}

Consider two phases $P$ and $Q$ that coexist in the temperature range $[T_{\rm min}, T_{\rm max}]$ and have a critical temperature at $T_c$. Additionally, suppose the model has a discrete symmetry, which transforms $P \to P'$ and $Q \to Q'$. Within the interval $[T_{\rm min}, T_c]$, suppose that phases $P$ and $P'$ are false vacuums, such as shown in \cref{fig:discrete symmetries}. We can obtain actions for the following possible transitions: 
\begin{align}
P\to Q,~P'\to Q,~P\to Q'~\text{and}~P'\to Q'.
\end{align}
According to the discrete symmetry, 
\begin{align}
S(P\to Q) &= S(P'\to Q'),\\
S(P'\to Q) &= S(P\to Q'),
\end{align}
but it is possible that 
\begin{align}
S(P\to Q) \neq S(P\to Q').
\end{align}
These two actions determine whether the Universe undergoes a transition from $P \to Q$ or $P \to Q'$. In \cref{fig:discrete symmetries}, the transition will clearly choose the path from $P \to Q$. 

In \code{PhaseTracer2}, we identify which of these transitions may still be unique after accounting for the discrete symmetry and then compute actions for all the unique transitions at the temperature
\begin{align} \label{eq:T_unique_actions}
T = T_{\rm min} + 0.9 \times( T_c- T_{\rm min} )
\end{align}
and consider the transition with the smallest action value to be the one that could have actually occurred in the model's cosmological history.

The factor $0.9$ in \cref{eq:T_unique_actions}  was chosen so that the temperature lies close to $T_c$ but not excessively so, as the bubble profile near $T_c$ exhibits a thin wall, which is challenging to calculate accurately. Conversely, near $T_{\rm min}$, the smallest action, corresponding $P\to Q$, is more straightforward to compute due to its correspondence with a thick wall. However, for other transitions to the symmetric minimum, $P\to Q'$ calculating the action may pose difficulties. $P$ evaporates at $T_{\rm min}$, so does the barrier between $P$ and $Q$. Thus the tunnelling path of $P \to Q'$ may go through the vicinity of $P'$, i.e.\ $P \to \text{near $P'$} \to Q'$. The path deformation method starts by guessing a straight tunnelling path $P \to Q'$, making it challenging to converge onto the correct path $P \to \text{near $P'$} \to Q'$
and prone to generating errors.
Simultaneously transitioning to more than one minimum and the formation of domain walls are not considered in this context. 

\subsection{Applications: Nucleation, Completion and duration of the phase transition}
\label{sec:Bounce_applications}

Extending \PT so that it can compute the bounce action makes it possible to use \PT in a number of important applications, including some that can be done directly within \PT.

As shown in \cref{Eq:Nuc_rate_zero_temp,Eq:Nuc_rate_finite_temp,A_est_zero_temp,A_est_finite_temp} the bounce action can be used to estimate the nucleation rate, $\Gamma/V$.  The bounce action is divergent at the critical temperature, such that the nucleation rate is zero initially. As such, the decay is not expected to occur immediately but as the universe cools the nucleation rate increases from zero and bubbles start to form. The nucleation temperature, $T_n$, is defined as the temperature at which the expected number of bubbles nucleated per Hubble horizon, $N(T)$, is one: $N(T_n) \sim 1$.   This gives a rough indication of when bubbles are starting to form and has been widely used in estimates of the baryon asymmetry of the universe and gravitational wave signatures in the literature.  

For a Hubble volume of order $H^{-3}$, the number of bubbles nucleated at temperature $T$ is given by \cite{Athron:2022mmm, PhysRevD.45.3415, Guo_Sinha_Vagie_White_2021}
\begin{equation}
    N(T) = \int_{T_C}^T dT' \frac{\Gamma(T') P_f(T')}{T' H^4(T') },
    \label{eq:NucleatedBubbles}
\end{equation}
where $P_f(T)$ is the false vacuum fraction at temperature $T$ (see Ref.~\cite{Athron:2023xlk} and references therein for a discussion of this and details on how to calculate it). Direct calculation of $N(T)$ is cumbersome, involving an evaluation of the bounce action and a sequence of nested integrals. For sufficiently fast transitions the integration can be simplified by approximating $P_f(T') \approx 1$ inside \cref{eq:NucleatedBubbles}. This is a good estimate in those cases because many small bubbles are nucleated and the nucleation temperature will then be reached while $P_f(T') \approx 1$ remains true.

Furthermore there exist many approximations to $T_n$ that avoid this integration altogether, see Ref.~\cite{Athron:2022mmm} for a discussion and comparison of these.  All of these approximations allow the nucleation temperature to be computed using the the bounce action available in \PT with relatively simple algorithms. By assuming $A \sim T^4$, the simplest approximation (see Appendix B of Ref.~\cite{Redondo_2009} for one way to obtain this), is
\begin{equation}\label{eq:simple_tn_condition}
    \frac{S_3(T_n)}{T_n} \sim 140.
\end{equation}
Using a bounce calculator to determine $S_3(T)$
allows \cref{eq:simple_tn_condition} to be used in conjunction with a root-finding method to find $T_n$. Although this is a very rough approximation, computing it is a simple application of the bounce action and allows direct comparisons with \code{CosmoTransitions}~\cite{wainwright2012cosmotransitions}) where it is used.   

Therefore as an illustration of the use of this, \code{PhaseTracer2} calculates the nucleation temperature using the method \code{get\_Tnuc} of the \code{TransitionFinder} class. 
This begins at the critical temperature and evaluates $S_3(T)/T - 140$, stepping down through temperature at an increment defined by \code{Tnuc\_step}, which by default is set to $1\gev$. If a temperature $T_1$ is found such that $S_3(T)/T - 140 < 0$, by \cref{eq:simple_tn_condition} we assume the transition nucleates. The nucleation temperature is then found by using a bisection-based root finding method, using $T_1$ as an initial guess. Instructions for activating this estimate of the nucleation temperature in the code are given in \cref{sec:running_pt2} and its use is illustrated in the following examples
\begin{itemize}
    \item \code{example/run\_1D\_test\_model.cpp}
    \item \code{example/run\_2D\_test\_model.cpp}
\end{itemize}

However it is important to note that as well as \cref{eq:simple_tn_condition} being a rough approximation of the nucleation temperature $T_n$, the nucleation temperature is not a fundamental property of the phase transition as the transition can complete without nucleation, and conversely nucleation can occur without phase transition completion \cite{Athron:2022mmm}. Furthermore it is is often a poor proxy for the percolation temperature ($T_p$) \cite{Athron:2022mmm}, which is a better choice for evaluating the GW spectrum. Using $T_n$ instead of $T_p$ can lead to orders of magnitude errors in GW predictions \cite{Athron:2023rfq}. To compute $T_p$ and the completion temperature, $T_f$, the false vacuum fraction must be computed.

Furthermore the approach described here only checks nucleation for an individual transition, but its possible that the false vacuum in a considered transition is not reached because an earlier transition does or does not complete in time, i.e.\ one should really algorithmically determine the full cosmological phase history.  These calculations are performed in \code{TransitionSolver} which uses \PT to identify all possible transitions and then determines the true cosmological phase history for the user input model and parameters.  \code{TransitionSolver} uses the bounce action to estimate the nucleation rate and it tests which paths complete using a bounce solver, while it uses elegant methods from graph theory to find the true history efficiently.   We refer the reader to the forthcoming manual \cite{Athron:2024tbc} for usage of \code{TransitionSolver}, but eager users may already try it out, see \cref{sec:TS}, albeit without documentation and at their own risk.

The bounce action can also be used to obtain a rough estimate of the duration of the phase transition, as the reciprocal of
\begin{equation}\label{eq:timescale_tn_condition}
    \beta(T) = \frac{d}{dt} \bigg ( \frac{S_3}{T} \bigg ) = T H(T) \frac{d}{dT} \bigg ( \frac{S_3}{T} \bigg ).
\end{equation}
Here $\beta$ is really the coefficient of exponential nucleation rate $\Gamma(t) = \Gamma(t_*) \exp[\beta (t-t_*)]$ which has been obtained by Taylor expanding the action about $t_*$ and truncating at first order. The $t_*$ is some characteristic time scale during the transition, often taken to be the time of nucleation.    

The nucleation temperature can also be used to estimate the level of supercooling 
via a supercooling parameter~\cite{Athron:2022mmm}
\begin{equation}
    \delta_{sc, n} = \frac{T_c - T_n}{T_c}.
\end{equation}
If this is too large then $\beta$ will not be a useful quantity to characterise the phase transition as the assumption of an exponential nucleation rate will not be good and $\beta$ may even be negative.
While $\beta$ should only be used for fast transitions, it often appears in fits to the gravitational wave amplitude, where it usually enters after substitution from some length scale that is controlled in the simulations.  This has the advantage of being much easier to compute for a specific model and we utilise this in \cref{sec:GravWaves}. However, see Ref.~\cite{Athron:2023rfq} for a discussion of the error associated with this approximation.

\section{Gravitational Wave Spectrum}\label{sec:GravWaves}

In a radiation-dominated Universe, there are three sources of GWs from a phase transition: bubble collisions, sound waves, and magnetohydrodynamic (MHD) turbulence.  After bubble nucleation, the pressure difference between the interior and exterior of the bubbles drives their expansion, ultimately leading to bubble collisions. These collisions disrupt the spherical symmetry of the bubbles, creating nonzero anisotropic stress, which serves as a source of GWs. Simultaneously, the collisions inject energy into the surrounding plasma, perturbing the plasma that initially possesses a high Reynolds number and inducing turbulence, thus generating anisotropic stress. Moreover, the expansion of the bubbles itself induces motion in the surrounding plasma and generates distinct plasma velocity profiles based on the bubble wall velocity, thereby contributing to the GWs.

In \code{PhaseTracer2} we have included routines to compute the GW spectrum from all three sources. These are usually estimated from expressions written as functions   of so-called thermal parameters that capture model dependence coming from the specific potential.  In \PT we use the inverse time-scale $\beta$ (see \cref{eq:timescale_tn_condition}), the trace anomaly definition of $\alpha$ (see  \cref{sec:alpha}) and the bubble wall velocity $v_w$ as thermal parameters. All of these in principle should be computed for the specific potential being considered, but reliable determinations of the bubble wall velocity outside of time consuming simulations are lacking currently,\footnote{See e.g. Refs.~\cite{Wang:2024slx,Tian:2024ysd} for an approach using fast Higgsless simulations \cite{Jinno_2021, Jinno:2022mie} to determine these things from the potential, which is complimentary to the approaches taken here, but is currently too slow for the purposes of \PT.} so we treat $v_w$ as an input parameter as is a common practice in these calculations.  Furthermore when determining the gravitational wave spectrum with \PT alone we do so at the nucleation temperature, computing it with a simple heuristic estimate as described in \cref{sec:Bounce_applications}. 

We note that this approach leads to significant uncertainties in fast transitions, as shown in Ref.~\cite{Athron:2023rfq} and is not suitable for slow transitions, i.e.\ when there is large supercooling.  To go beyond this one needs to use \code{TransitionSolver} in combination with \PT.  With \code{TransitionSolver} the percolation temperature can be computed, appropriate checks for true percolation and completion of the transition can be made and more sophisticated approaches for incorporating the model dependence of the effective potential are used.  The calculations in \code{TransitionSolver} are essentially as described in the Appendix of Ref.~\cite{Athron:2023mer}, but proper documentation for \code{TransitionSolver} will be released in a forthcoming manual \cite{Athron:2024tbc}.  However users may try the current beta-version by installing \code{TransitionSolver} as described in \cref{sec:TS}.

In the following we instead document the routines for calculating the gravitational wave spectra natively in \PT.  The main difference to the routines in \code{TransitionSolver} being that they are written in terms of the thermal parameters that we can calculate straightforwardly in \PT and evaluated at the nucleation temperature and we do not check for true percolation or completion. 

\subsection{Bubble Collisions}

For bubble collisions, the envelope approximation is a commonly used method where bubbles are treated as infinitely thin, and in regions where bubbles overlap, the bubble walls are completely ignored. This approach was first introduced in Ref.~\cite{PhysRevD.45.4514} and has since been widely adopted to reduce computational demands on the lattice~\cite{PhysRevD.45.4514,PhysRevD.47.4372,Huber:2008hg}. Based on Ref.~\cite{Jinno:2016vai}, we use the fitting formula for the gravitational wave spectrum under the envelope approximation
\begin{equation} \label{eq: omega_col}
\begin{split}
\Omega^{\rm env}_{\rm col} h^2  ={}& 1.67 \times 10^{-5} \, \Delta \, \left(\frac{100}{g_{*}}\right)^{\frac{1}{3}}\left(\frac{H_{*}}{\beta}\right)^2 \\
& \times \left(\frac{\kappa_\phi \alpha}{1+\alpha}\right)^2  
\, S_{\rm env}(f / f_{\rm env}),
\end{split}
\end{equation}
with
\begin{equation}
\Delta = \frac{0.48v_w^3}{1+5.3v_w^2+5v_w^4},
\end{equation}
where $v_w$ represents the velocity of the bubble wall. The spectral shape,
\begin{equation}
S_{\rm env}(r) = \left(0.064 r^{-3} + 0.456 r^{-1} 
+ 0.48 r\right)^{-1},
\end{equation}
with peak frequency at
\begin{align}
\frac{f_{\rm env}}{1\,\mu\mathrm{Hz}}&=16.5 \left(\frac{f_{*}}{\beta}\right)\left(\frac{g_{*}}{100}\right)^{\frac{1}{6}} \left(\frac{\beta}{H_{*}}\right) \left(\frac{T_{\star}}{100\gev}\right) ,\\
\frac{f_{*}}{\beta} &= \frac{0.35}{1+0.069v_w + 0.69v_w^4},
\end{align}
where $H_{*}$ denotes the Hubble constant at the reference temperature $T_{*}$, which is typically selected as either the nucleation temperature or the percolation temperature. 
Lastly, the efficiency factor $\kappa_{\phi}$ determines how much of the vacuum energy is transformed into kinetic energy of the bulk fluid instead of reheating the plasma inside the bubble. It could be calculated by the fitting formula~\cite{Kamionkowski:1993fg} 
\begin{equation}
    \kappa_{\phi} = \frac{1}{1 + 0.715 \alpha} \left[0.715\alpha + \frac{4}{27} \sqrt{\frac{3 \alpha}{2}}\right].
\end{equation}
\subsection{Turbulence}
For the turbulence contribution, Ref.~\cite{Caprini:2009yp, Binetruy:2012ze} obtained the frequently adopted fitting formulas
\begin{equation} \label{eq: omega_turb}
\begin{split}
\Omega_{\mathrm{turb}} h^2={}& 3.35 \times 10^{-4} \, v_w \, \left(\frac{H_*}{\beta}\right)\left(\frac{\kappa_{\mathrm{turb}} \alpha}{1+\alpha}\right)^{3 / 2}
\\
& \times
\left(\frac{100}{g_*}\right)^{1 / 3} 
\frac{r^3}{(1 + r)^{11 / 3}\left(1+8 \pi f / H_0\right)},
\end{split}
\end{equation}
where $r = f / f_{\mathrm{turb}}$ with peak frequency
\begin{equation}
f_{\mathrm{turb}}=27 \frac{1}{v_w}\left(\frac{\beta}{H_n}\right)\left(\frac{T_*}{100\gev}\right)\left(\frac{g_*}{100}\right)^{1 / 6} ~\mu\mathrm{Hz},
\end{equation}
the red-shifted Hubble rate at GW generation $H_0$ is 
\begin{equation}
H_0=16.5 \, \left(\frac{g_{*}}{100}\right)^{\frac{1}{6}} \left(\frac{T_{\star}}{100\gev}\right)~\mu\mathrm{Hz},
\end{equation}
and the $\kappa_{\mathrm{turb}}$ is the energy fraction transferred to MHD turbulence and can vary between 5\% to 10\% of $\kappa_{\rm sw}$~\cite{PhysRevD.92.123009}, which we will talk about later.

\subsection{Sound Waves}
From numerical simulations we anticipate that sound waves could  act as the primary mechanism for generating GWs from a first-order phase transition~\cite{Hindmarsh:2013xza}. There are  fitting formulas for the acoustic contribution~\cite{PhysRevD.92.123009, Hindmarsh:2017gnf, PhysRevD.101.089902,Caprini:2019egz}\footnote{Note that in the original paper~\cite{PhysRevD.92.123009} this equation contains typos; see the erratum~\cite{PhysRevD.101.089902}.}
\begin{equation} \label{eq: omega_sw}
\begin{split}
    \Omega_{\mathrm{sw}} h^2={}& 2.061 \, F_{\rm gw, 0} \, \Gamma^2 \, \Bar{U}_{f}^{4} \, S_{\rm sw}(f) \, \tilde\Omega_{\rm gw} \\
    &\times \min(H_{*}R_{*}/\Bar{U}_{f}, 1) \, \left(H_{*}R_{*}\right)h^2,
\end{split}
\end{equation}
where
\begin{align}
    F_{\rm gw,0} &= 3.57 \times 10^{-5} \left(\frac{100}{g_{*}}\right)^{\frac{1}{3}},  \\
    S_{\rm sw}(f) &= \left(\frac{f}{f_{\rm sw}}\right)^3 \left(\frac{7}{4 + 3 \left(f/f_{\rm sw}\right)^2}\right)^{\frac{7}{2}},  \\
    \frac{f_{\rm sw}}{1\,\mu\mathrm{Hz}} &= 2.6 \left(\frac{z_p}{10}\right) \left(\frac{T_{*}}{100\mathrm{GeV}}\right) \left(\frac{g_{*}}{100}\right)^{\frac{1}{6}} \left(\frac{1}{H_{*}R_{*}}\right).
\end{align}
In \cref{eq: omega_sw}, $\Gamma$ is the ratio of enthalpy to the energy density for the fluid, which is often set to $4/3$ for the early Universe. The quantities $z_p \sim 10$  and $\tilde\Omega_{\rm gw} \sim 0.012$ are determined from simulations. $R_{*}$ is the mean bubble separation and can be roughly calculated by 
\begin{equation}
    R_{*} = \left(8 \pi\right)^{1/3} v_w / \beta.
\end{equation}
The term $\Bar{U}_{f}$ represents the enthalpy-weighted root mean square of the fluid velocity. By the definition of these quantities, we can relate the combination $\Gamma \Bar{U}_{f}^{2}$ to the kinetic energy fraction $K$ as~\cite{Caprini:2019egz, Hindmarsh:2017gnf}
\begin{equation}
K = \frac{\kappa_{\rm sw} \alpha}{1 + \alpha} = \Gamma \Bar{U}_{f}^{2},
\end{equation}
where $\kappa_{\rm sw}$ is the efficiency factor, which can be obtained through the energy budget analysis~\cite{Espinosa:2010hh}. The function $\min(H_{*}R_{*}/\Bar{U}_{f}, 1)$ accounts the shock formation time effect (this suppression factor is also used by the web-based tool \href{https://www.ptplot.org/ptplot/}{\code{PTPlot}}). Theoretically, this suppression factor can also be derived within the sound shell model~\cite{Guo:2020grp, RoperPol:2023dzg, Guo:2024kfk}, and the corresponding gravitational wave spectrum has a fitting formula~\cite{Guo:2024gmu}. To meet the needs of rapid parameter scanning, \code{PhaseTracer2} employs the following fitting formula based on the bag model and constant sound velocity $c_s= 1/\sqrt{3}$ to estimate the efficiency factor $\kappa_{\rm sw}$~\cite{Espinosa:2010hh}
\begin{align}
    \kappa_{\rm sw} = 
\begin{cases} 
  \frac{c_s^{11/5}\kappa_A \kappa_B}{\left(c_s^{11/5}  - v_w^{11/5}\right)\kappa_B  + v_wc_s^{6/5}\kappa_A}, & \!\!\!v_w \le  c_s \\[2mm]
  \kappa_B + (v_w-c_s) \delta \kappa + \frac{\left(v_w-c_s\right)^3}{(v_J - c_s)^3}l_{\kappa}, & \!\!\!c_s < v_w < v_J  \\[2mm] 
  \frac{\left(v_J-1\right)^3v_J^{5/2}v_w^{-5/2}\kappa_C \kappa_D}{\left[\left(v_J-1\right)^3 - \left(v_w-1\right)^3\right] v_J^{5/2} \kappa_C + \left(v_w-1\right)^3\kappa_D}, & \!\!\! v_J \le  v_w
\end{cases}
\end{align}
where
\begin{align}
    \kappa_A &\simeq v_w^{6/5} \frac{6.9  \alpha}{1.36-0.037\sqrt{\alpha} + \alpha},  \\
    \kappa_B  &\simeq \frac{\alpha^{2/5}}{0.017 + \left(0.997 + \alpha\right)^{2/5}}, \\
    \kappa_C &\simeq \frac{\sqrt{\alpha}}{0.135 + \sqrt{0.98 + \alpha}}, \\
    \kappa_D &\simeq \frac{\alpha}{0.73 + 0.083\sqrt{\alpha} + \alpha}, \\
    \delta \kappa &\simeq -0.9{\rm log}\frac{\sqrt{\alpha}}{1 + \sqrt{\alpha}},  \\
    l_{\kappa} &\simeq \kappa_C - \kappa_B - \left(v_J - c_s\right) \delta \kappa, \\
    v_J &= \frac{1}{1+\alpha} \left(c_s + \sqrt{\alpha^2 + \frac{2 \alpha}{3}}\right).
\end{align}

\subsection{Combined GW spectrum and signal to noise ratio in \PT}
After collecting all the ingredients, \code{PhaseTracer2} calculates the final GW spectrum
\begin{equation}
    \Omega_{\rm gw}h^2 = \Omega_{\rm col}^{\rm env}h^2 + \Omega_{\rm sw}h^2 + \Omega_{\rm turb} h^2,
\end{equation}
using the \code{GravWaveCalculator} class. Its method \code{calc\_spectrums} will automatically obtain the information of $\alpha$, $\beta/H$, $v_w$, and $T_{*}$ based on the given \code{TransitionFinder} object and use them to calculate the corresponding GW spectrums and the corresponding signal-to-noise ratio for LISA~\cite{Smith:2019wny, Gowling:2022pzb}.  
The signal-to-noise ratio for Taiji~\cite{10.1093/nsr/nwx116} and Tianqin~\cite{TianQin:2015yph} will be included at a later stage. At high temperatures, the contribution of bubble collisions to the gravitational wave spectrum is significantly smaller than that of sound waves. Therefore, \code{PhaseTracer2} by default neglects the bubble contribution when $T_{*} > 10\gev$.

To get the GW specturms in \code{PhaseTracer2}, consult the following examples
\begin{itemize}
    \item \code{example/run\_1D\_test\_model.cpp}
    \item \code{example/run\_2D\_test\_model.cpp}
\end{itemize}
or \cref{sec:running_pt2}.

\section{\code{PhaseTracer2} User Guide}\label{sec:UserGuide}

\begin{figure*}[htbp]
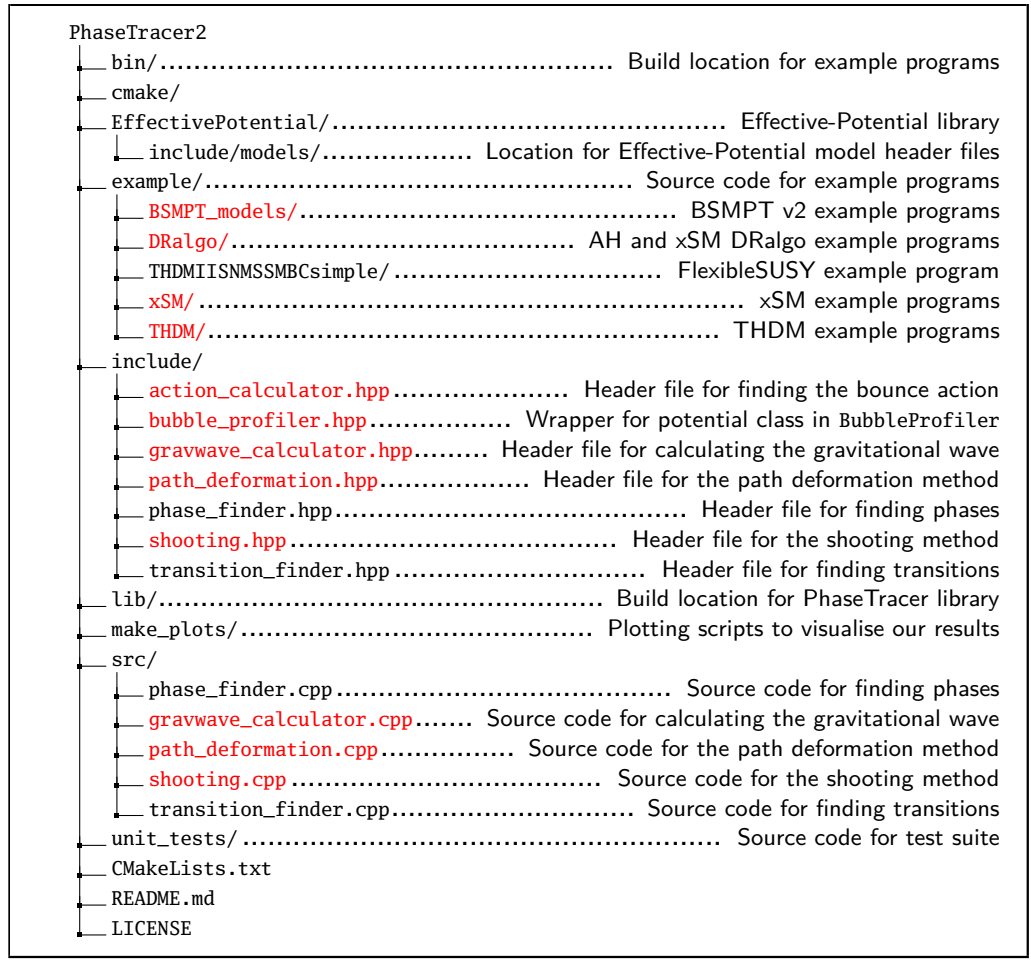

\centering
\begin{minipage}[t]{0.75\linewidth}
  \begin{mdframed}[linewidth=0.3mm]
    \dirtree{%
      .1 \code{PhaseTracer2}.
      .2 bin/\DTcomment{Build location for example programs}.
      .2 cmake/.
      .2 EffectivePotential/\DTcomment{Effective-Potential library}.
      .3 include/models/\DTcomment{Location for Effective-Potential model header files}.
      .2 example/\DTcomment{Source code for example programs}.
      .3 \textcolor{red}{BSMPT\_models/}\DTcomment{BSMPT v2 example programs}.
      .3 \textcolor{red}{DRalgo/}\DTcomment{AH and xSM DRalgo example programs}.
      .3 THDMIISNMSSMBCsimple/\DTcomment{FlexibleSUSY example program}.
      .3 \textcolor{red}{xSM/}\DTcomment{xSM example programs}.
      .3 \textcolor{red}{THDM/}\DTcomment{THDM example programs}.
      .2 include/.
      .3 \textcolor{red}{action\_calculator.hpp}\DTcomment{Header file for finding the bounce action}.
      .3 \textcolor{red}{bubble\_profiler.hpp}\DTcomment{Wrapper for potential class in \code{BubbleProfiler}}.
      .3 \textcolor{red}{gravwave\_calculator.hpp}\DTcomment{Header file for calculating the  gravitational wave}.
      .3 \textcolor{red}{path\_deformation.hpp}\DTcomment{Header file for the path deformation method}.
      .3 phase\_finder.hpp\DTcomment{Header file for finding phases}.
      .3 \textcolor{red}{shooting.hpp}\DTcomment{Header file for the shooting method}.
      .3 transition\_finder.hpp\DTcomment{Header file for finding transitions}.
      .2 lib/\DTcomment{Build location for PhaseTracer library}.
      .2 make\_plots/\DTcomment{Plotting scripts to visualise our results}.
      .2 src/.
      .3 phase\_finder.cpp\DTcomment{Source code for finding phases}.
      .3 \textcolor{red}{gravwave\_calculator.cpp}\DTcomment{Source code for calculating the  gravitational wave}.
      .3 \textcolor{red}{path\_deformation.cpp}\DTcomment{Source code for the path deformation method}.
      .3 \textcolor{red}{shooting.cpp}\DTcomment{Source code for the shooting method}.
      .3 transition\_finder.cpp\DTcomment{Source code for finding transitions}.
      .2 unit\_tests/\DTcomment{Source code for test suite}.
      .2 CMakeLists.txt.
      .2 README.md.
      .2 LICENSE.
      }
  \end{mdframed}
  \end{minipage}
  \medskip
  \caption{The file structure of \code{PhaseTracer2}, wherein only significant files have been included. Files new to \code{PhaseTracer2} have been highlighted in red.}
  \label{fig:PT2filestructure}
\end{figure*}

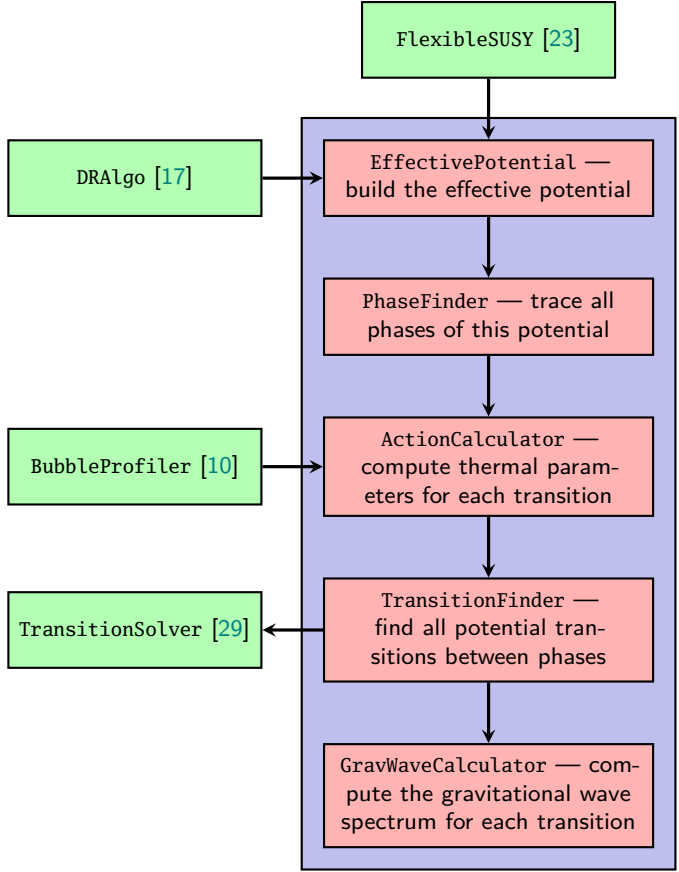
\begin{figure}
\centering
\begin{tikzpicture}[node distance=0.8cm,font=\sf]
%
%
\node (a) [box] {\code{EffectivePotential} --- build the effective potential};
\node (b) [box, below=of a] {\code{PhaseFinder} --- trace all phases of this potential};
\node (c) [box, below=of b] {\code{ActionCalculator} --- compute thermal parameters for each transition};
\node (d) [box, below=of c] {\code{TransitionFinder} --- find all potential transitions between phases};
\node (e) [box, below=of d] {\code{GravWaveCalculator} --- compute the gravitational wave spectrum for each transition};
%
%
\node (bp) [depbox, left=of c] {\code{BubbleProfiler}~\cite{Athron_2019}};
\node (dralgo) [depbox, left=of a] {\code{DRAlgo}~\cite{Ekstedt_2023}};
\node (fs) [depbox, above=of a] {\code{FlexibleSUSY}~\cite{ATHRON2018145}};
\node (ts) [depbox, left=of d] {\code{TransitionSolver}~\cite{Athron:2024tbc}};
%
%
\draw [arrow] (bp) -- (c);
\draw [arrow] (dralgo) -- (a);
\draw [arrow] (fs) -- (a);
\draw [arrow] (d) -- (ts);
\draw [arrow] (a) -- (b);
\draw [arrow] (b) -- (c);
\draw [arrow] (c) -- (d);
\draw [arrow] (d) -- (e);
\begin{scope}[on background layer]
    \node [pt, fit=(a) (e)] {};
\end{scope}
\end{tikzpicture}
\medskip
\caption{Outline of \code{PhaseTracer} calculations (purple). The external codes (green) are optional.}
\label{fig:pt2_stages}
\end{figure}

In \cref{fig:PT2filestructure}, the updated file structure of \code{PhaseTracer2} has been provided, wherein new additions detailed in previous \cref{sec:EffPot,sec:TransProp,sec:GravWaves} have been outlined in red. 
With these additions included, the physics problem addressed by \code{PhaseTracer2} can now be broken up into the stages in \cref{fig:pt2_stages}.

\renewcommand{\DTstylecomment}{~\sf}

The remainder of this section is now dedicated to a user guide for \code{PhaseTracer2}. We will start with a preliminary remark on the implementation of new models, then move on to explain each part of the code and how it addresses the aforementioned physics problems.

\subsection{Creating New Models}

For a given model, creation of the effective potential used by \code{PhaseTracer2} is done using the \code{EffectivePotential} library, which can be found within the base package directory. This can be used to implement both the tree-level potential via overwriting the virtual \code{Potential} class, or the one-loop potential via overwriting the virtual \code{OneLoopPotential} class. Whilst \code{PhaseTracer2} features an array of new model types, the core functionality of the \code{EffectivePotential} library has not changed.\footnote{The exception here is the new functionality required to incorporate the dimensionally reduced potential from \code{DRalgo}. Further details on this can be found in \cref{sec:DRalgoExp}.} To that end, the reader is referred to Section 5.1 of the original \code{PhaseTracer} manual for more details on model implementation. Additionally, the model header files for the examples in \cref{sec:Examples} contain valuable guidance on how the original functionality can be repurposed to accommodate the many new model types. 

\subsection{Building \code{PhaseTracer2}} \label{sec:building_pt2}

In addition to the header files and source files for running the model, a \code{CMakeLists.txt} is also required, with the following content:
\begin{lstlisting}
add_executable(run_model_name run_model_name.cpp)
\end{lstlisting}
where \code{run\_model\_name} should be replaced with the user-defined name of the executable. If these files are placed within a folder inside the \code{example} directory, \code{CMake} will automatically detect them, and the project can be compiled by executing:
\begin{lstlisting}[language=bash]
$ cd PhaseTracer
$ mkdir build
$ cd build
$ cmake ..
$ make <model_name>
\end{lstlisting}
Then, the built executable can be run using
\begin{lstlisting}[language=bash]
$ ./bin/<run_model_name>
\end{lstlisting}

\subsection{Running \code{PhaseTracer2}} \label{sec:running_pt2}

\newcommand{\modelName}{\itshape\textcolor{orange}{<ModelName>}}

Once the user has chosen or created a model file using the \code{EffectivePotential} library, one can proceed to call this model within a program to utilise \code{PhaseTracer2}. Here we provide a step-by-step instruction manual to perform this task. Example code will be provided for each step, such that this code could be strung together to provide a cohesive and functioning program. Here \code{\modelName} should be replaced with one's chosen header file for a model. All example outputs shown in this section are taken from the dimensionally reduced scalar singlet model, found in \cref{sec:dim_reduced_ssm}.

To initialise the example program, all relevant files must be included. This can be performed with the following code:
\begin{lstlisting}
#include <iostream>
#include "models/|\modelName|.hpp"
#include "phasetracer.hpp"
\end{lstlisting}

Then, the effective potential must be created using the following code:
\begin{lstlisting}
// create effective potential object
EffectivePotential::ModelName potential;
\end{lstlisting}
The remaining functionality of \code{PhaseTracer2} is spread across four classes: \code{PhaseFinder}, \code{TransitionFinder}, \code{ActionCalculator}, and \code{GravWaveCalculator}. These handle the jobs of phase finding, transition finding, bounce action evaluation, and gravitational wave spectrum calculation respectively. Each class must be initialised, then the respective functionality can be accessed using the \code{find\_phases()}, \code{find\_transitions()}, and
\code{calc\_spectrums()} methods. The \code{ActionCalculator} class contains a single public method, \code{get\_action}, and is intended to be used in the initialisation of the \code{TransitionFinder} object. 

In the following sections, we detail the use of each class, including example code, relevant data structures, and methods and attributes.

\subsubsection{\code{PhaseFinder}}

\code{PhaseFinder} performs the task of phase tracing the provided effective potential. It is created and utilised with the following example code:
\begin{lstlisting}
// create phase finder object
PhaseTracer::PhaseFinder pf(potential)
// find phases
pf.find_phases();
// print the results
std::cout << pf;
\end{lstlisting}
Printing the result produces the following output:
\begin{lstlisting}[keywordstyle=\ttfamily]
found 2 phases

=== phase key = 0 ===
Maximum temperature = 1000
Minimum temperature = 73.1614
Field at tmax = [2.11456e-05, 2.12428e-05]
Field at tmin = [-0.00122037, -101.362]
Potential at tmax = 0.000393378
Potential at tmin = -2.88489e+07
Ended at tmax = Reached tstop
Ended at tmin = Hessian was not positive definite

=== phase key = 1 ===
Maximum temperature = 132.463
Minimum temperature = 15
Field at tmax = [-68.2468, 5.27385e-05]
Field at tmin = [-208.372, 2.57285e-05]
Potential at tmax = -737568
Potential at tmin = -7.7718e+07
Ended at tmax = Jump in fields indicated end of phase
Ended at tmin = Reached tstop
\end{lstlisting}
This summary includes every phase found, numbered by the key \code{phase key}. For each phase, the relevant information provided includes the temperature range this phase exists, the field configuration at either temperature boundary, the corresponding value of the potential, and lastly the reason the phase ends.

\begin{table*}[t]
\centering
\begin{tabularx}{0.9\textwidth}{llX}
  \toprule
  Type & Name & Description \\
  \midrule
  \code{Phase}\\
  \midrule
  \code{size\_t} & \code{key} & Reference key for each phase. \\
  \code{std::vector<Eigen::VectorXd>} & \code{X} & Location of minima $x_{\text{min}}(T)$. \\
  \code{std::vector<double>} & \code{T} & Temperature of phase, in ascending order. \\
  \code{std::vector<Eigen::VectorXd>} & \code{dXdT} & Temperature gradient of minima $dx_{\text{min}}/dT$. \\
  \code{std::vector<double>} & \code{V} & Potential $V(x_{\text{min}})$ throughout the temperature range. \\
  \code{bool} & \code{redundant} & Specifies the redundancy of the phase, used to account for discrete symmetries. \\
  \code{phase\_end\_descriptor} & \code{end\_low} & Reason for the phase ending at the lower temperature limit. \\
  \code{phase\_end\_descriptor} & \code{end\_high} & Reason for the phase ending at the upper temperature limit. \\
  \midrule
  \code{Transition}\\
  \midrule
  \code{size\_t} & \code{key} & Reference key for the transition. \\
  \code{double} & \code{TC} & Critical temperature of the transition. \\
  \code{double} & \code{TN} & Nucleation temperature of the transition. \\
  \code{Phase} & \code{true\_phase} & Phase for the true vacuum. \\
  \code{Phase} & \code{false\_phase} & Phase for the false vacuum. \\
  \code{Eigen::VectorXd} & \code{true\_vacuum} & True vacuum at $T_c$. \\
  \code{Eigen::VectorXd} & \code{false\_vacuum} & False vacuum at $T_c$. \\
  \code{Eigen::VectorXd} & \code{true\_vacuum\_TN} & True vacuum at $T_n$. \\
  \code{Eigen::VectorXd} & \code{false\_vacuum\_TN} & False vacuum at $T_n$. \\
  \code{double} & \code{gamma} & The ratio $v(T_c)/T_c$, where $v$ is the vev. \\
  \code{std::vector<bool>} & \code{changed} & Status of which fields participated (true) or did not participate (false) in the transition. \\
  \code{double} & \code{delta\_potential} & Change in the potential between each minima at the critical temperature. \\
  \midrule
  \code{Spectrum}\\
  \midrule
  \code{double} & \code{Tref} & Reference temperature used to calculate the spectrum. \\
  \code{double} & \code{alpha} & Strength factor of the transition. \\
  \code{double} & \code{beta\_H} & Ratio of characteristic time scale to the Hubble parameter, $\beta(T_*)/H(T_*)$. \\
  \code{double} & \code{peak\_frequency} & Peak frequency of the spectrum, in Hz. \\
  \code{double} & \code{peak\_amplitude} & Peak amplitude of the spectrum. \\
  \code{std::vector<double>} & \code{frequency} & Frequency values used to calculate the spectrum. \\
  \code{std::vector<double>} & \code{sound\_wave} & Sound wave contribution to $\Omega h^2$. \\
  \code{std::vector<double>} & \code{turbulence} & MHD turbulence contribution to $\Omega h^2$\\
  \code{std::vector<double>} & \code{bubble\_collision} & Bubble wall collision contribution to $\Omega h^2$ \\
  \code{std::vector<double>} & \code{total\_amplitude} & Total summed $\Omega h^2$ from the three above components.\\
  \code{std::vector<double>} & \code{SNR} & SNR evaluated at implemented experiments (presently only LISA) \\
  \bottomrule
\end{tabularx}
\caption{Data structure of the various \code{Phase}, \code{Transition}, and \code{Spectrum} objects.}
\label{tab:structs}
\end{table*}

This information is stored within the \code{Phase} object, which can be accessed using \code{pf.get\_phases()}. This creates a vector of these \code{Phase} objects, with one corresponding to each listed phase. These are structs that contain all the information provided above, and can be used to access individual information regarding any particular phase. A detailed account of this struct can be found in \cref{tab:structs}. All vector quantities are listed in ascending temperature order (i.e., \code{T.front()} is the minimum temperature). For example, if one wanted to print the start and end temperature of the second EWSB phase, this information can be accessed from the \code{p} object as follows:
\begin{lstlisting}[language=c++]
// extract phases
auto p = pf.get_phases();
double max_temp = p[1].T.back();
double min_temp = p[1].T.front();
\end{lstlisting}

The \code{PhaseFinder} class includes various settings that modify and control its behavior, such as the maximum and minimum temperature bounds. These can be modified using setters (e.g.~\code{set\_\{PROPERTY\}}) and viewed using getters (e.g.~\code{get\_\{PROPERTY\}}). For the sake of brevity, the reader is referred to Section 5.2.2 of the original \code{PhaseTracer} paper, and Table 1 therein for a summary of these settings.

\subsubsection{\code{ActionCalculator}}

The \code{ActionCalculator} class contains functionality for the evaluation of the bounce action, including finding the nucleation temperature. The methods available in this object are not intended to be used publicly within the example program, but instead they are employed by the transition finding class. To that end, the user simply needs to initialise the \code{ActionCalculator} object for later use. This can be done with the following code:
\begin{lstlisting}[language=c++]
// create action calculator object
PhaseTracer::ActionCalculator ac(potential);
\end{lstlisting}
Whilst not needed in the program, the user should take note of the \code{get\_action} method. This takes three inputs: two objects of type \code{Eigen::VectorXd} corresponding to two vacuums, and a \code{double} corresponding to the temperature. It then returns the action evaluated between these two vacuums:
\begin{lstlisting}[language=c++]
double action = ac.get_action(true_vacuum, false_vacuum, T);
\end{lstlisting}

In \cref{tab:acSettings} the associated settings can be found, and again be read or modified using the associated getter or setter commands. Of note here are those used to specify whether the bounce action is evaluated using the interface with \code{BubbleProfiler}, or whether the in-house path deformation method is employed. By default, if \code{BubbleProfiler} is installed both options will be employed. \code{get\_action} will then return the minimum output of both calculations.

\begin{table*}[t]
\centering
\begin{tabularx}{0.9\textwidth}{llX}
  \toprule
  Setting & Default value & Description \\
  \midrule
  \code{num\_dims} & \code{3} & The dimension of the $O(n)$-symmetric field configuration. \\
   \midrule
  \code{BP\_use\_perturbative} & \code{false} & Specifies BP to always use the perturbative method, as opposed to utilising the shooting method for single field potentials. \\
  \code{BP\_initial\_step\_size} & \code{1e-2} & Step size for solving ODEs. For the adaptive step size of the shooting method, this sets the initial value. \\
  \code{BP\_interpolation\_points\_fraction} & \code{1} & Fraction of total grid sites used when building cubic spline interpolations. \\
   \midrule
  \code{PD\_xtol} & \code{1e-4} & In the Shooting algorithm, the precision of field values after taking the logarithm. \\
  \code{PD\_phitol} & \code{1e-4} & The cut off for finding the initial conditions for integration in the Shooting algorithm. \\
  \code{PD\_thin\_cutoff} & \code{1e2} & The cut off for finding the initial conditions for integration in the Shooting algorithm. \\
  \code{PD\_npoints} & \code{500} & Number of points to return in the profile of the Shooting algorithm. \\
  \code{PD\_rmin} & \code{1e-4} & The smallest starting radius in the Shooting algorithm. \\
  \code{PD\_rmax} & \code{1e4} & The maximum allowed integration distance in the Shooting algorithm. \\
  \code{PD\_max\_iter} & \code{100} & In the Shooting algorithm, the maximum number of points to be positioned during the integration process. \\
  \code{PD\_nb} & \code{10} & Number of basis splines to use in the path deformation algorithm. \\
  \code{PD\_kb} & \code{3} & Order of basis splines in the path deformation algorithm. \\
  \code{PD\_save\_all\_steps} & \code{false} & Flag to save path in each step of the path deformation algorithm. \\
  \code{PD\_v2min} & \code{0} &The smallest the square of ${\rm d}\phi/{\rm d} r$ is allowed to be in the path deformation algorithm.\\
  \code{PD\_path\_maxiter} & \code{100} &Number of samples to take along the path to create the spline
   interpolation functions.\\
  \code{PD\_step\_maxiter} & \code{500} & Maximum number of steps to take in a deformation.\\
  \code{PD\_path\_maxiter} & \code{20} & Maximum number of allowed deformation iterations.\\
  \code{PD\_extend\_to\_minima} & \code{true} & Flag to extend the path to minimums.\\
  \bottomrule
\end{tabularx}
\caption{Adjustable settings controlling the \code{ActionCalculator} class.}
\label{tab:acSettings}
\end{table*}

\subsubsection{\code{TransitionFinder}}

Once all phases have been founded, \code{TransitionFinder} performs the task of finding any transitions between them. As such, it must be initialised with both the \code{PhaseFinder} and \code{ActionCalculator} objects. This can be done with the following example code:
\begin{lstlisting}
// create transition finder object
PhaseTracer::TransitionFinder tf(pf, ac)
// find transitions
tf.find_transitions();
// print the results
std::cout << tf;
\end{lstlisting}
Printing the result  produces the following output:
\begin{lstlisting}[keywordstyle=\ttfamily]
found 1 transition

=== transition from phase 0 to phase 1 ===
changed = [true, true]
TC = 125.388
false vacuum (TC) = [-1.42526e-05, -58.5449]
true vacuum (TC) = [-100.341, 5.99233e-06]
gamma (TC) = 0.926498
delta potential (TC) = 0.00594705
TN = 118.444
false vacuum (TN) = [4.31038e-05, -67.3572]
true vacuum (TN) = [-122.489, -4.08245e-05]
\end{lstlisting}
This summary includes every transition found, numbered by \code{key}. For each transition, the relevant information provided includes which fields participate in the transition, the critical temperature, the nucleation temperature, the true and false vacuums at the critical temperature, the change in the potential at $T_c$, and lastly $\gamma(T_c) = v(T_c)/T_c$, where $v(T_c)$ is the VEV.

As with the phases, this information is contained within the \code{Transition} struct, and can be obtained using \code{get\_transitions()} which returns a vector of these objects. An account of this object can again be found within \cref{tab:structs}. For example, if one wanted to print the critical and nucleation temperatures for the transition above, these can be accessed as follows:
\begin{lstlisting}[language=c++]
// extract transitions
auto t = tf.get_transitions();
double crit_temp = t[0].TC;
double nuc_temp = t[0].TN;
\end{lstlisting}

As with the \code{PhaseFinder} class, the various settings that control the behaviour of the \code{TransitionFinder} class can be found within Section 5.2.2 and Table 2 of the original \code{PhaseTracer} paper.

\subsubsection{\code{GravWaveCalculator}}

Once all transitions are found, \code{GravWaveCalculator} can be used to calculate their corresponding gravitational wave spectrum. As such, this must be initialised with the \code{TransitionFinder} object. This is achieved with the following code:
\begin{lstlisting}[language=c++]
// create grav wave calculator object
PhaseTracer::GravWaveCalculator gc(tf);
// find all spectrums
gc.calc_spectrums();
// print the results
std::cout << gc;
\end{lstlisting}
\Cref{tab:gcSettings} contains the associated settings that control the \code{GravWaveCalculator} class.

Printing the \code{GravWaveCalculator} using \code{std::cout << gc;} produces the following output:
\begin{lstlisting}
found 1 spectrum

=== gravitational wave spectrum generated at T = 118.444 ===
alpha = 0.00157305
beta over H = 2950.73
peak frequency = 0.103761
peak amplitude = 3.02827e-22
signal to noise ratio for LISA = 4.89729e-12

=== total gravitational wave spectrum ===
peak frequency = 0.103761
peak amplitude = 3.02827e-22
\end{lstlisting}
This output can be broken into two parts. In the first part, for each detected transition the output shows the thermal parameters, peak frequency and amplitude, and  associated signal-to-noise ratios. The second part of the output shows the total spectrum (found by summing each of the constituent transitions). For the total sum, only the peak amplitude and frequency are shown.

This information can be extracted using the \code{get\_spectrums()} method. This extracts a vector of \code{Spectrum} objects. A summary of this object is also provided in \cref{tab:structs}. Note by default the spectrum is calculated at the nucleation temperature. For example, the amplitude of the sound wave contribution to the spectrum from the first transition can be accessed by
\begin{lstlisting}
std::vector<Spectrum> spec = gc.get_spectrums();
std::vector<double> amp = spec[0].sound_wave;
\end{lstlisting}
The frequencies corresponding to these amplitudes can be accessed by
\begin{lstlisting}
std::vector<double> freq = spec[0].frequency;
\end{lstlisting}

\begin{table*}[t]
\centering
\begin{tabularx}{0.9 \textwidth}{llX}
  \toprule
  Setting & Default value & Description \\
  \midrule
  \code{dof} & \code{106.75} & Relativistic degrees of freedom used for the radiation energy density. \\
  \code{vw} & \code{0.3} & Bubble wall velocity. \\
  \code{epsilon} & \code{0.1} & Ratio of turbulence and sound wave efficient factors, $\kappa$. \\
  \code{run\_time\_LISA} & \code{3} & Effective run time at LISA (years). \\
  \code{T\_obs\_LISA} & $\mathtt{3 \times 365.25 \times 24 \times 60 \times 60}$ & Acquisition time for LISA (seconds). \\
  \code{SNR\_f\_min} & \code{1e-5} & Lower integration bound for SNR integration. \\
  \code{SNR\_f\_max} & \code{1e-1} & Upper integration bound for SNR integration. \\
  \code{min\_frequency} & \code{1e-4} & Lower frequency bound on gravitational wave spectrum. \\
  \code{max\_frequency} & \code{1e5} & Upper frequency bound on gravitational wave spectrum. \\
  \code{num\_frequency} & \code{500} & Number of data points used in frequency discretisation. \\
  \code{T\_threshold\_bubble\_collision} & \code{10} & Upper bound for bubble collision calculation. If $T_*$ is above this, the bubble collision spectrum will not be calculated. \\
  \code{h\_dVdT} & \code{1e-2} & Step size for numerical evaluation of $dV/dT$. \\
  \code{h\_dSdT} & \code{1e-1} & Step size for numerical evaluation of $dS/dT$. \\
  \code{n\_dSdT} & \code{5} & Number of points sampled for numerical evaluation of $dS/dT$. \\
  \bottomrule
\end{tabularx}
\caption{Adjustable settings controlling the \code{GravWaveCalculator} class.}
\label{tab:gcSettings}
\end{table*}

\subsubsection{Plotting Results and Additional Functions}

In addition to printing the results as above, \code{PhaseTracer2} provides various functions for visualising results. These include the following functions: \code{phase\_plotter}, \code{potential\_line\_plotter}, \code{potential\_plotter}, and \code{spectrum\_plotter}. Details on the first three of these, which are carried over from \code{PhaseTracer1}, can be found in Section 5.3 of the original \code{PhaseTracer} publication. \code{spectrum\_plotter} is used to plot the gravitational wave spectrum, and is called with the following syntax:
\begin{lstlisting}
PhaseTracer::spectrum_plotter(gc, "model_name");
\end{lstlisting}
Here \code{gc} is the \code{GravWaveCalculator} object, and prior to being used the spectrums must be calculated using \code{gc.calc\_spectrums()}. \code{"model\_name"} is a string used in the naming of the resulting plots. That is, the plots will be saved under the name: \code{"model\_name\_GW\_spectrum\_i"}, where \code{i} runs over the individual spectrums. An example of the created plots can be found in \cref{fig:3dEFT_spec}.

Additionally, we provide a means of extracting the numerical values of the spectrum. The function \code{write\_spectrum\_to\_text} takes an individual spectrum (e.g. \code{gw[i]}) and writes it to a text file, or writes all the spectrums into separate files. It is called with a similar syntax:
\begin{lstlisting}
  gc.write_spectrum_to_text(i, "file_name.txt");
  gc.write_spectrum_to_text("file_name.txt");
\end{lstlisting} 
The first line writes a single spectrum to a file named \code{"file\_name.txt"} in CSV format. The second line write each spectrum to a file named \code{"i\_file\_name.txt"}.
The first column of this file contains frequency in Hz, the second column shows the corresponding total gravitational wave spectrum $\Omega_{\rm gw}h^2$, the third column indicates the contribution from sound waves $\Omega_{\rm sw}h^2$, the fourth column is the contribution from turbulence $\Omega_{\rm turb}h^2$, and the last column provides the contribution from bubble collisions $\Omega_{\rm col}^{\rm env}h^2$.

\section{Examples} \label{sec:Examples}

\subsection{Polynomial potentials} \label{sec:ExamplePolynomial}

We demonstrate the bounce action capabilities of \code{PhaseTracer2} by considering the following single-field potential:
\begin{equation}\label{eq:example_potential}
    V(\phi) = E \mu^4 \left [ \frac{-4\alpha+3}{2} \phi^2 -\phi^3 +\alpha \phi^4 \right ],
\end{equation}
taken from \code{BubbleProfiler}. This is equivalent to an arbitrary fourth-order potential, where under re-scaling we have eliminated any additional free parameters. This model can be run using:
\begin{lstlisting}
$ ./bin/run_BP_scale E alpha scale
\end{lstlisting}
where \code{E}, \code{alpha}, and \code{scale} correspond to $E$, $\alpha$, and $\mu$ respectively in \cref{eq:example_potential}.  For example, running with
\begin{equation}
    \{E, \alpha,\mu\} = \{ 1., 0.6, 200. \}
\end{equation}
will return $S = 1544.68$. In addition to this, we also consider the following 2D test model taken from section 5.1 of the \code{CosmoTransitions} manual \cite{wainwright2012cosmotransitions}:
\begin{equation}
    V(\phi_1,\phi_2) = (\phi_1^2 + \phi_2^2)(1.8(\phi_1 - 1)^2 + 0.2(\phi_2-1)^2 - 0.4).
\end{equation}
This can be run using
\begin{lstlisting}
$ ./bin/run_BP_2d
\end{lstlisting}
which will return $S = 12.9423$. In this program we have determined the minima of the potential using the method \code{get\_minima\_at\_t\_low()} from the \code{PhaseFinder} class. We have then utilised the \code{get\_action()} method to calculate the bounce action between these two minima.

The valued returned in the above example has utilised the default functionality of \code{get\_action()}, which is to run through each available action calculator and return the minimum value. In this case, we have only provided the path deformation algorithm. In each section below, the results are calculated using this method without the interface with \code{BubbleProfiler}.

\subsection{Two-dimensional Test Model} \label{sec:Example2D}

As with the first version of \pt, we implement the 2d test model found in section 5.2 of the \code{CosmoTransitions} manual, where a detailed description of this model can be found \cite{wainwright2012cosmotransitions}. To summarise, the model contains two scalar fields, $\phi_1$ and $\phi_2$, mixing through a quadratic $\phi_1 \phi_2$ portal term. Additionally, a heavy scalar boson is included to break the degeneracy of the minima, allowing for a first-order transition. The tree-level potential is given as 
\begin{equation}
    V_0(\phi_1,\phi_2) = \frac{1}{8}\frac{m_1^2}{v^2} (\phi_1^2-v^2)+\frac{1}{8}\frac{m_2^2}{v^2} (\phi_2^2-v^2) - \mu^2\phi_1\phi_2,
\end{equation}
where $m_1=120\gev$, $m_2=50\gev$, $\mu = 25\gev$ and $v = 246\gev$ by default. 
This model can be run using 
\begin{lstlisting}
$ ./bin/run_2D_test_model
\end{lstlisting}

\begin{table*}[t]
\centering
\begin{tabular}{llllll}
  \toprule
  & $T_C$ & $T_N$ & False VEV & True VEV &  Time (s) \\
  \midrule
  \code{CosmoTransitions} & 109.4 & 84.24 & $[231.1,-136.8]$ & $[286.4,382.2]$ & 2.93\\
  \midrule
  \code{PhaseTracer} & 109.4 & 84.24 & $[231.1,-136.8]$ & $[286.4,382.2]$ & 0.61 \\
  \bottomrule
\end{tabular}
\caption{Comparison of results and elapsed time between \code{PhaseTracer} and \code{CosmoTransitions} for the 2d test model in Section 7.2, using a desktop with a Intel Core i7-12800H CPU $@$ 4.8GHz processor.
Note that \pt performs additional action calculations in order to ascertain the appropriate transition path. 
All dimensionful quantities are given in units of GeV.}
\label{tab:PtvsCt}
\end{table*}

\begin{figure}
    \centering
    \includegraphics[width=\linewidth]{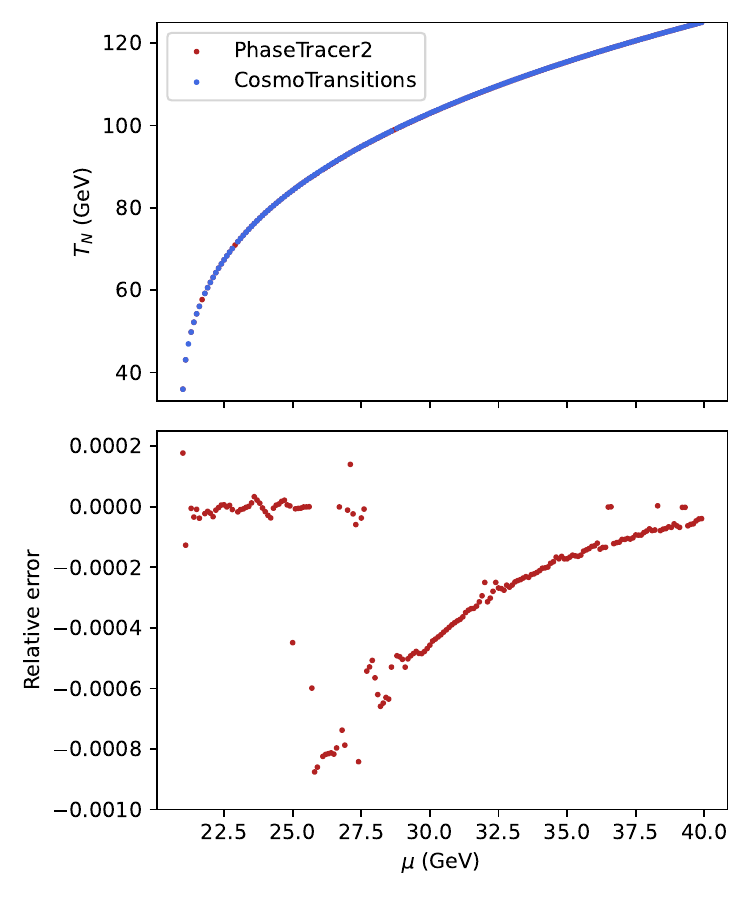}
    \caption{Comparison of $T_N$ and relative error between \code{PhaseTracer} and \code{CosmoTransitions} for the 2d test model. }
    \label{fig:TN_error}
\end{figure}

We demonstrated in \code{PhaseTracer1} that our code yields excellent agreement for the location of minima and the critical temperature, whilst performing these calculations approximately 100 times faster. Here, we further the calculation to include an evaluation of the nucleation temperature. In \cref{tab:PtvsCt} and \cref{fig:TN_error} we compare the updated performance of \code{PhaseTracer2}. 
We can see that the two tools yield consistent results across the entire scan region. The relative errors for $T_N$ are always smaller than 1\%. In the upper panel of \cref{fig:TN_error}, the red dots representing results from \code{PhaseTracer} are mostly covered by the blue dots from \code{CosmoTransitions}. However, there are three red dots that are not covered because \code{CosmoTransitions} encounters errors for those points.  There is also one point where \code{PhaseTracer} encounters an error.
The scan using \code{CosmoTransitions} takes 9 minutes and 32 seconds, while the scan using \code{PhaseTracer} takes 1 minute and 57 seconds. 
This example has a $\ztwo$ discrete symmetry. As explained in Section \cref{sec:discrete symmetries}, it necessitates the computation of additional actions within \code{PhaseTracer} compared to \code{CosmoTransitions}. Therefore, the improvement in the speed of calculating the nucleation temperature is limited.

\subsection{Dimensionally Reduced Abelian Higgs Model} \label{sec:DRalgo_ah}

In \cref{sec:DRalgoExp}, we provided detail on how to implement the 3DEFT obtained from \code{DRalgo} into our \code{PhaseTracer2} code. In this section, as well as in \cref{sec:dim_reduced_ssm}, we provide examples of this implementation. Whilst the reader is likely familiar with the Abelian Higgs model, we provide a short description below. This model features a single $U(1)$ gauge interaction, with a complex Higgs field $\phi$:
\begin{equation}
    \mathcal{L} = - \frac{1}{4} F_{\mu \nu} F^{\mu \nu} + | D_{\mu} \phi |^2 - V(\phi),
\end{equation}
where the scalar potential is
\begin{equation}
    V(\phi) = \mu^2 |\phi|^2 + \lambda (|\phi|^2)^2.
\end{equation}
Local gauge invariance is ensured by the covariant derivative:
\begin{equation}
    D_{\mu} \phi = \partial_{\mu} \phi - i g A_{\mu} \phi
\end{equation}
for gauge boson $A_{\mu}$. Provided $\mu^2 < 0$, the potential above obtains the non-zero VEV:
\begin{equation}
    \langle \phi \rangle = \sqrt{ \frac{\mu^2}{2 \lambda}},
\end{equation}
which we take to lie in the purely real direction: $\langle \phi \rangle = h /\sqrt{2}$. We then exchange $\mu$ for the mass of our physical Higgs field $h$, $m$, using:
\begin{equation}
    \mu^2 = - \frac{1}{2} m^2.
\end{equation}
Thus, we have three free parameters $\{ m, g^2, \lambda \}$. 

The source code for this model can be found in the file:
\begin{lstlisting}
$ PhaseTracer/examples/DRalgo/DRalgo_ah.hpp
\end{lstlisting}
and can be run by using:
\begin{lstlisting}
$ ./bin/run_DRalgo_ah <m> <gsq> <lam> <potential> <matching>
\end{lstlisting}
The first three arguments correspond to our model parameters. The additional arguments \code{potential} and \code{running} are flags to specify the accuracy to which we evaluate the potential and matching respectively. For \code{potential}, 0 corresponds to tree-level, 1 to one-loop, and 2 to the 2-loop potential. Similarly, an argument of 0 for \code{matching} performs the dimensional reduction at the hard scale to LO, whilst 1 matches to NLO, and lastly an argument of 2 performs the dimensional reduction at the soft scale, such that the 3d parameters gain additional NLO corrections due to integrating out temporal gauge bosons (note that the soft dimensional reduction is formally NLO). 

In \cref{fig:AHplot} we show this potential for the benchmark point $\{m, g^2, \lambda \} = \{ 125\gev, 0.42, 0.128 \}$. We have utilised the minimal approach, setting \code{potential = matching = 1}. These also correspond to the default options for these settings. Within the program itself, these options are treated as properties of the potential, and as such can be modified using getter and setters, e.g.:
\begin{lstlisting}[language=c++]
model.set_potential_flag(1);
model.set_matching_flag(1);
\end{lstlisting}
\begin{figure}
    \centering
    \includegraphics[width=\linewidth]{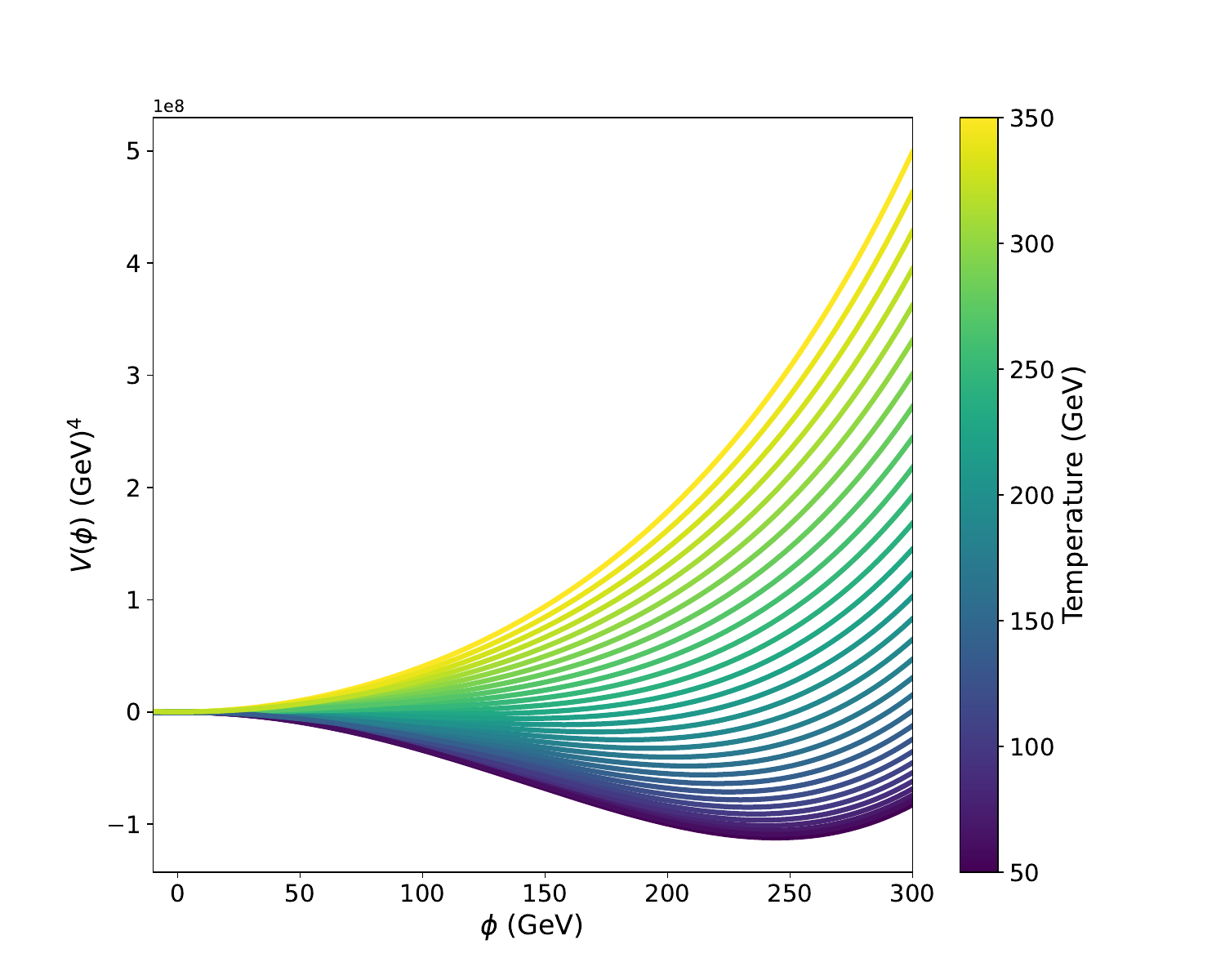}
    \caption{The potential of the Abelian Higgs model at various temperatures.}
    \label{fig:AHplot}
\end{figure}

\subsection{\texorpdfstring{$\ztwo$}{Z2} Scalar Singlet Model}\label{sec:ztwo_ssm}

In \code{PhaseTracer2} we include the capability to investigate different renormalisation schemes, high temperature expansions, and different renormalisation scales. Here we demonstrate these for the $\ztwo$ scalar singlet model. In particular, we look at the high temperature expansion, the \msbar one loop potential, and OS-like one loop potential, and the PRM $\hbar$-expansion. An explanation of these schemes can be found in \cref{sec:EffPot}. The scalar singlet extension is the addition of a real gauge singlet to the SM, described by the scalar potential:
\begin{equation}
    \begin{split}
    V(H,S) ={} & \mu_H^2 |H^{\dagger} H| + \mu_S S^2 \\ & + \lambda_H |H^{\dagger} H|^2 + \lambda_S S^4 + \lambda_{HS} |H^{\dagger} H| S^2.
    \end{split}
\end{equation}
Here we have the VEV alignments $H = (0, h)^T/\sqrt{2}$ and $S = s/\sqrt{2}$. In the electroweak symmetry breaking vacuum $(h, s) = (246, 0)\gev$, this model predicts two massive scalar bosons, with physical masses given by:
\begin{equation}
    \begin{split}
        m_h^2 & = 2 \lambda_H v^2 \\
        m_s^2 & = \mu_s^2 + \lambda_{HS} v^2 / 2\\
    \end{split}
\end{equation}
This is the simplest extension of the SM that permits both a strongly first-order EWSB phase transition, and a $125\gev$ Higgs mass. We refer the reader to Ref.~\cite{Vaskonen:2016yiu} for a more complete description of this model.

The various included variations of the xSM can be run using the following commands:
\begin{lstlisting}
$ ./bin/run_xSM_HT ms lambda_s lambda_hs
$ ./bin/run_xSM_OSlike ms lambda_s lambda_hs daisy_flag
$ ./bin/run_xSM_MSbar ms lambda_s lambda_hs Q xi daisy_flag use_1L_EWSB_in_0L_mass use_Goldstone_resum
$ ./bin/run_xSM_PRM  ms lambda_s lambda_hs Q xi daisy_flag use_1L_EWSB_in_0L_mass use_Goldstone_resum
\end{lstlisting}
Here \code{ms} is the physical scalar mass, whilst \code{lambda\_s} and \code{lambda\_hs} correspond to the scalar-self coupling and scalar-Higgs portal couplings. The setting \code{Q} refers to the MSbar renormalisation scale, and \code{xi} to the $R_{\xi}$ gauge parameter. Furthermore, \code{daisy\_flag} refers to the resummation scheme, with the values \code{0}, \code{1}, and \code{2} corresponding to no resummation, Parwani resummation, and Arnold-Espinosa resummation respectively. 
\begin{figure}
    \centering
    \includegraphics[width=\linewidth]{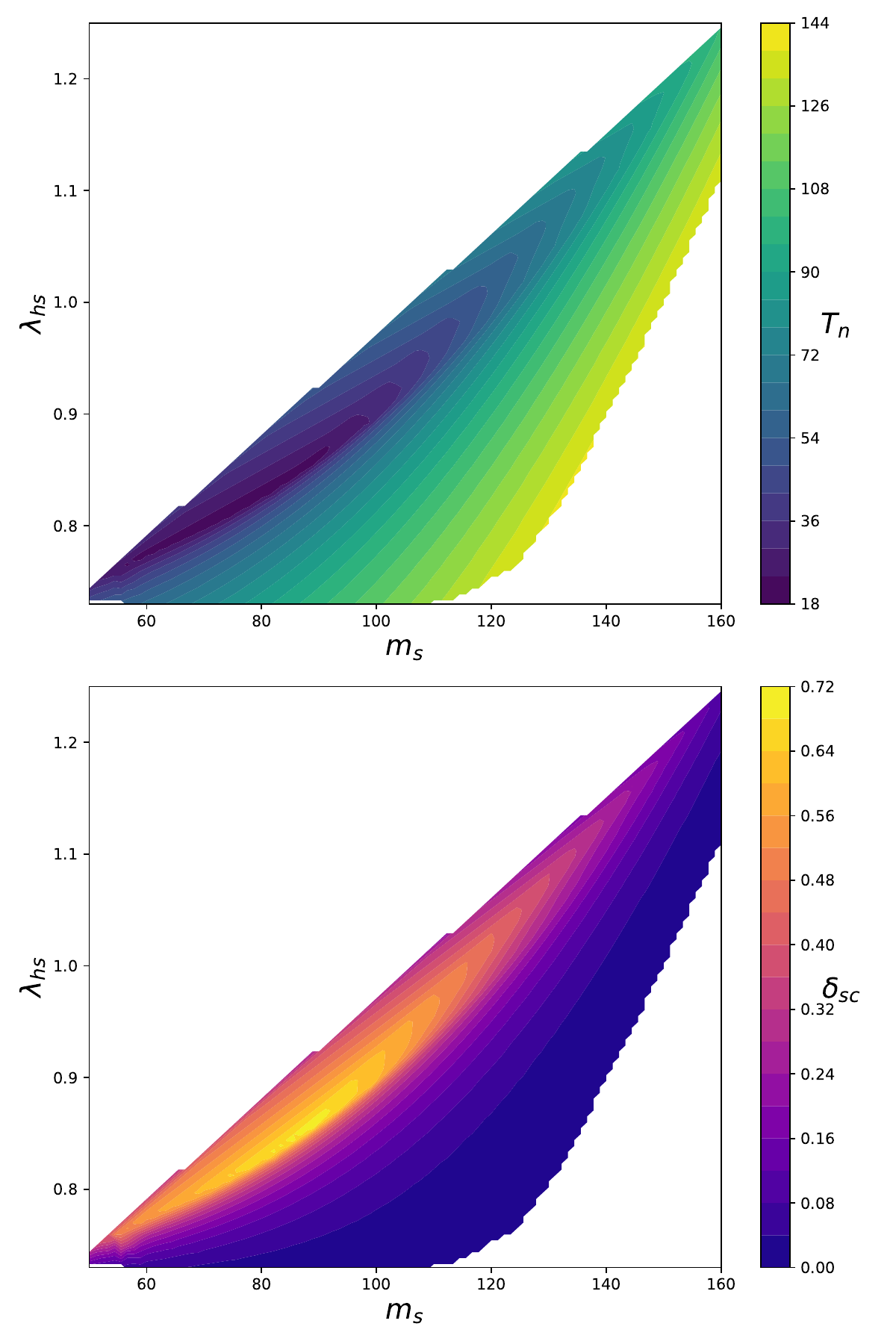}
    \caption{Top: The nucleation temperature of the $\mathbb{Z}_2$ symmetric scalar singlet model, calculated using the high-temperature expansion, as a function of physical scalar mass $m_s$ and Higgs-scalar portal coupling $\lambda_{HS}$. Bottom: The supercooling parameter $\delta$ for the same model. In both cases $\lambda_S = 1$.}
    \label{fig:HT_nuc}
\end{figure}

For the remainder of this section, we will utilise the high-temperature expansion to showcase the new functionality of \code{PhaseTracer2}. To begin,  we perform a scan of the nucleation temperature across the xSM parameter space. The scalar self-interaction has been fixed to $\lambda_S = 1$, and the physical scalar mass and scalar-Higgs portal coupling have been sampled across the range $0.01 < \lambda_{HS} < 1.5$ and $75\gev < m_s < 160\gev$. The top panel of in \cref{fig:HT_nuc} shows the results of this scan. In the lower panel, we further contextualise these results by using the nucleation temperature in conjunction with the critical temperature to evaluate the supercooling parameter $\delta_{sc}$:
\begin{equation}
    \delta_{sc} = \frac{T_c - T_n}{T_c},
\end{equation}
as utilised in Ref.~\cite{Athron:2022mmm}. 

\begin{figure}
    \centering
    \includegraphics[width=\linewidth]{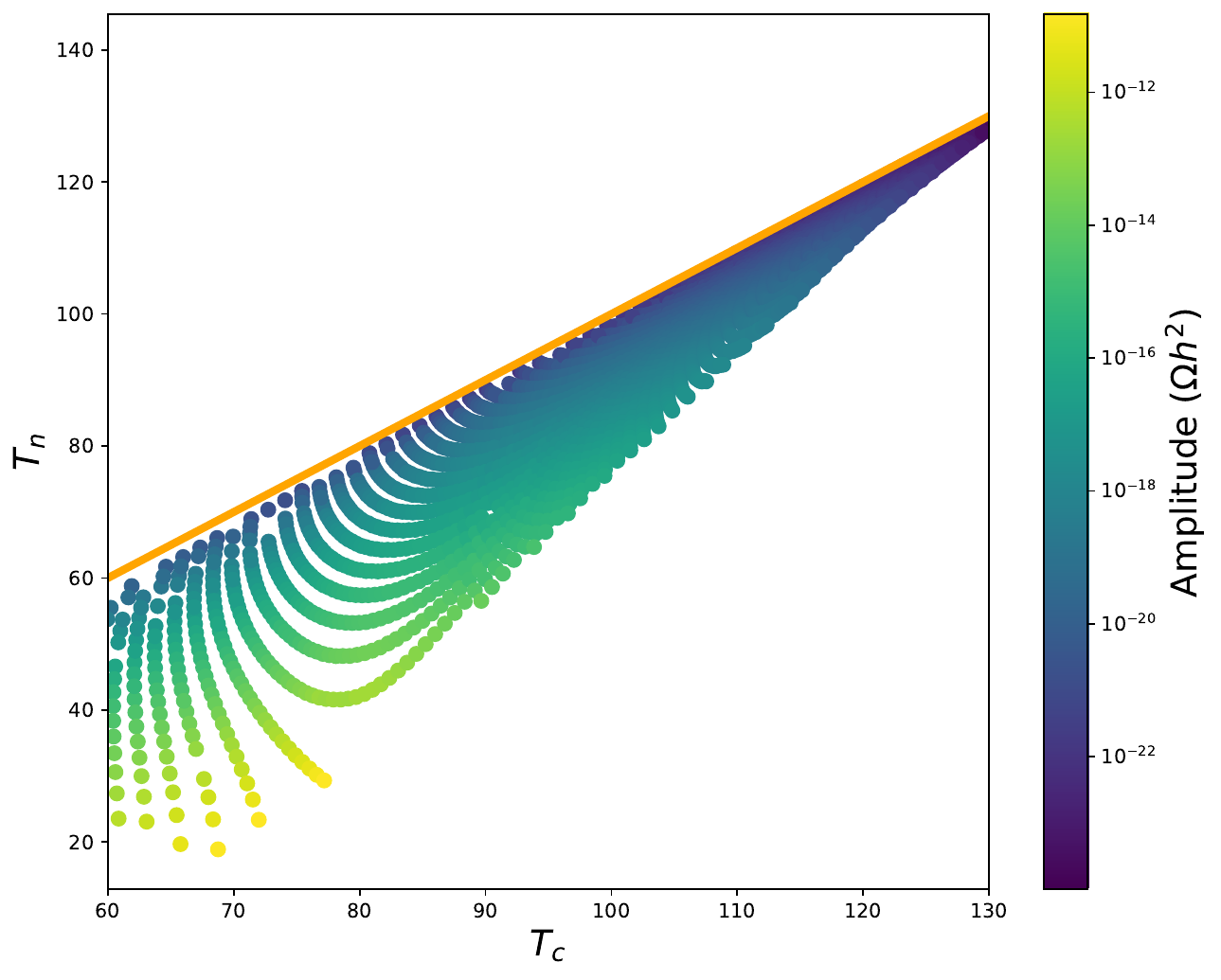}
    \caption{The peak gravitational wave amplitude, $\Omega h^2$, of the $\mathbb{Z}_2$ symmetric scalar singlet model, calculated using the high-temperature expansion, as a function of the critical and nucleation temperatures. Here $\lambda_S = 1$, $50 < m_{s} < 160\gev$, and $0.01 < \lambda_{HS} < 1.5$. The dotted orange line corresponds to the fast transition limit, $T_n \to T_c$.}
    \label{fig:HT_Tn_Tc}
\end{figure}

In \cref{fig:HT_Tn_Tc} we calculate the peak gravitational wave amplitude. We show these results as a function of both the nucleation and critical temperatures, where the orange dashed line indicates $T_n = T_c$, or the fast transition limit. As expected, the strongest gravitational wave signal is observed in regions with a large deviation from this line, corresponding to strong supercooling.

\begin{figure}
    \centering
    \includegraphics[width=\linewidth]{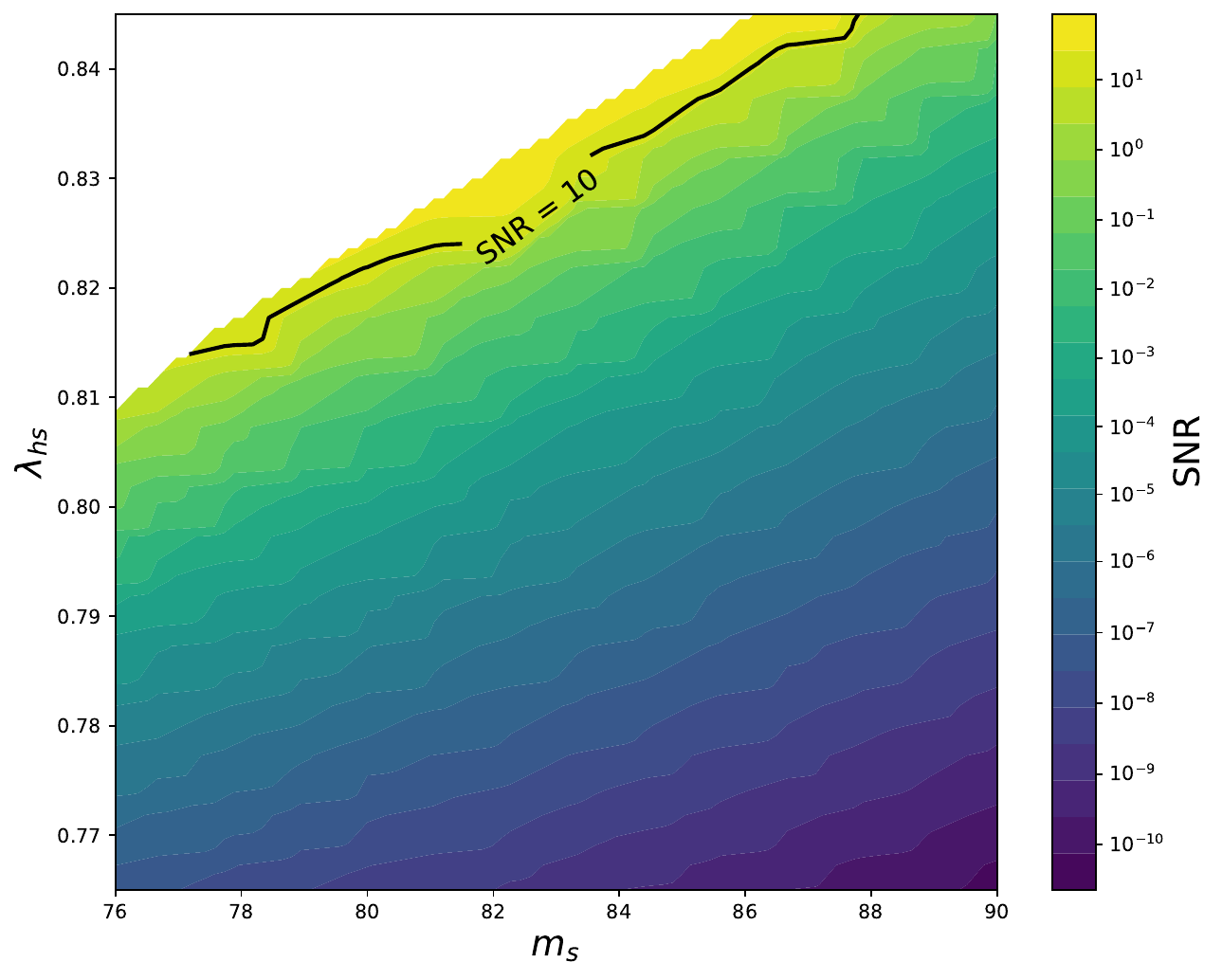}
    \caption{The SNR at LISA evaluated for transitions with the high temperature xSM. Results are shown over the limited range $76 < m_s < 90\gev$ and $0.765 < \lambda_{hs} < 0.845$, where again $\lambda_s = 1$. The black line shows the detectability threshold $\text{SNR} = 10$.}
    \label{fig:HT_SNR}
\end{figure}

In \cref{fig:HT_SNR}, the SNR to noise ratio has been evaluated from a limited sample of the parameter space, corresponding to regions with the strongest gravitational wave amplitude. The solid line on this plot indicates the threshold condition $\text{SNR} > 10$, deemed to mark the minimum requirement for detectability.

\begin{figure}
    \centering
    \includegraphics[width=\linewidth]{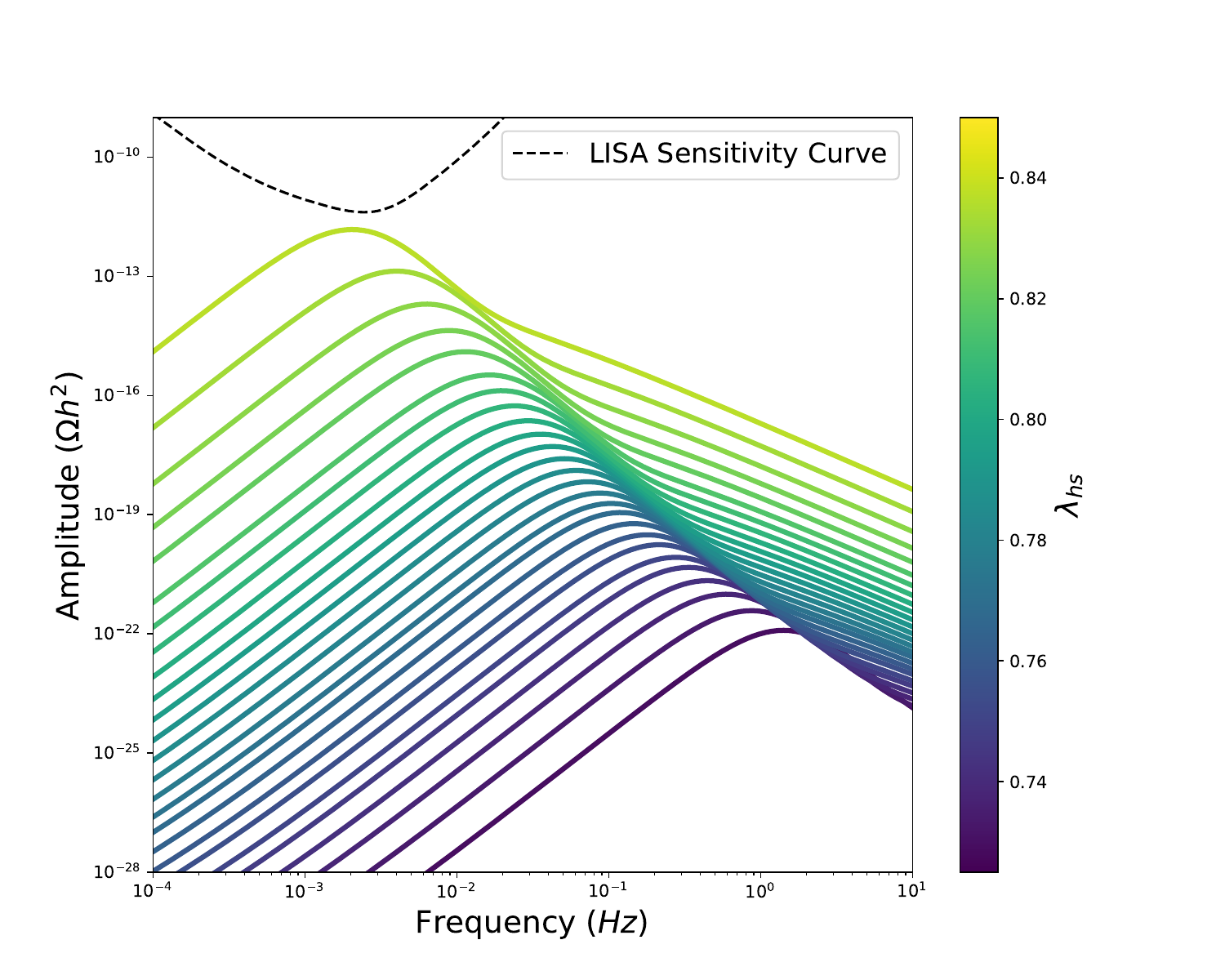}
    \caption{The gravitational wave spectrum, $\Omega h^2(f)$ calculated for the EWSB transition in $\mathbb{Z}_2$ symmetric scalar singlet model, utilising the high temperature expansion. Results have been calculated for the benchmark point $m_s = 83.7\gev$ and $\lambda_S = 1$. $\lambda_{HS}$ has been sampled in the range $0.725 < \lambda_{HS} < 0.85$. Shown in the dotted black line is the expected sensitivity curve of LISA, obtained using \href{https://www.gwplotter.com}{GWplotter}.}
    \label{fig:HT_spec}
\end{figure}

To conclude this section, in \cref{fig:HT_spec} we further restrict our parameter space to the benchmark parameters $m_s = 83.7\gev$ and $\lambda_S = 1$, and scan over the portal couplings $0.725 < \lambda_{hs} < 0.85$. This plot now shows the full gravitational wave spectrum, where a predictive LISA sensitivity curve has been included.

\subsection{Dimensionally Reduced Scalar Singlet Model} \label{sec:dim_reduced_ssm}

We previously demonstrated the implementation of dimensionally reduced theories by using the Abelian Higgs toy model. Here, we now showcase this framework for the scalar-singlet model defined in \cref{sec:ztwo_ssm}. As before, all relevant quantities from the dimensional reduction have been obtained with the use of \code{DRalgo}. This includes all 3D quantities, 4D beta functions, and the effective potential up to two-loop order. The relevant model header files can be found in:
\begin{lstlisting}
$ PhaseTracer/examples/DRalgo/DRalgo_xsm.hpp
$ PhaseTracer/examples/DRalgo/DRalgo_xsm_nnlo.hpp
\end{lstlisting}
The latter contains the expression for the two-loop correction to the effective potential which, due to the complexity of this expression, has been stored in a separate model file. Similar to the examples from \cref{sec:ztwo_ssm}, this model can be run using:
\begin{lstlisting}
$ ./bin/run_DRalgo_xSM <ms> <lambda_s> <lambda_hs>
\end{lstlisting}
These parameters have the same meaning as those used in \cref{sec:ztwo_ssm}. Alternatively, users wishing for more control can run the model as follows:
\begin{lstlisting}
$ ./bin/run_DRalgo_xSM <ms> <lambda_s> <lambda_hs> <running> <potential> <matching>
\end{lstlisting}
Here, \code{running} is a Boolean used to specified whether the quantities from the 4D parent theory are first run to the matching scale (defined as $\mu = \pi T$) prior to dimensional reduction. If $\code{running = false}$, all dimensionally reduced quantities will be calculated using the parameters defined at the input $\mu = 246$ GeV. As before, both \code{potential} and \code{matching} are accuracy flags for the effective potential and matching procedure respectively. 

\begin{figure}
    \centering
    \includegraphics[width=\linewidth]{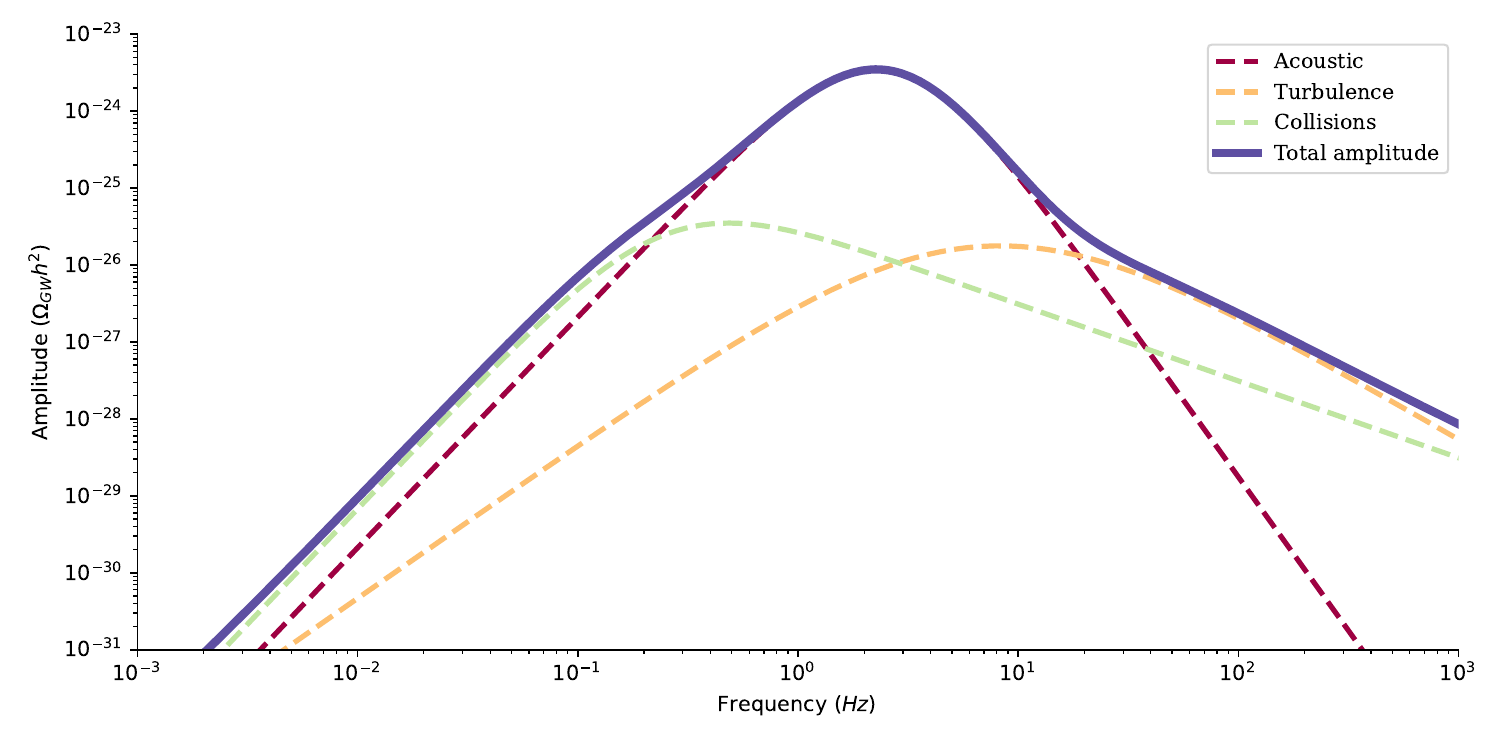}
    \caption{Gravitational wave spectrum for the transition in \cref{sec:dim_reduced_ssm}. The dotted lines signify the contribution from each source, whilst the solid line indicates the total summed spectrum. Here $T_n > 10\gev$, so by default the gravitational waves sourced from bubble collisions would not be computed. This plot was created utilising the \code{spectrum\_plotter} function.}
    \label{fig:3dEFT_spec}
\end{figure}

Unlike in the Abelian Higgs, the two-loop potential has been manually disabled, due its poor impact on compilation time. Users wishing to utilise the full two-loop potential must re-enable it by removing the comments in the definition of \code{DR\_xsm::V}, located within the model header file. For the default values of these settings, we again use the minimal approach:
\begin{lstlisting}
running = true
potential = 1
matching = 1
\end{lstlisting}
Once again, these can all be changed within the program itself using getter and setter options. We now investigate this model for the following benchmark point:
\begin{equation}
    \{m_s, \lambda_{s}, \lambda_{hs} \} = \{ 135\gev, 1, 1.05 \}
\end{equation}
We choose to employ RG evolution of all parent parameters, as well as performing the matching at NLO. Furthermore, we show these results initially using the tree-level potential. The reason for this is such that, up to corrections of order $\mathcal{O}(\lambda^2)$, this is equivalent to a RG improved HT expansion. Recall the example output in \cref{sec:running_pt2} was obtained using these inputs.
For this choice of parameters, we have the two step transition:
\begin{equation}
    (0, 0) \rightarrow (0, s) \rightarrow (h, 0),
\end{equation}
in which the first step is a crossover at tree-level, whilst the second transition is of first-order. In \cref{fig:3dEFT_spec} we show the potential for various temperatures, and \cref{fig:3dEFT_spec} shows the gravitational wave spectrum corresponding to the second step FOPT. Note that this plot was obtained using the built in \code{spectrum\_plotter} functionality new to \code{PhaseTracer2}.

\begin{figure}
    \centering
    \includegraphics[width=\linewidth]{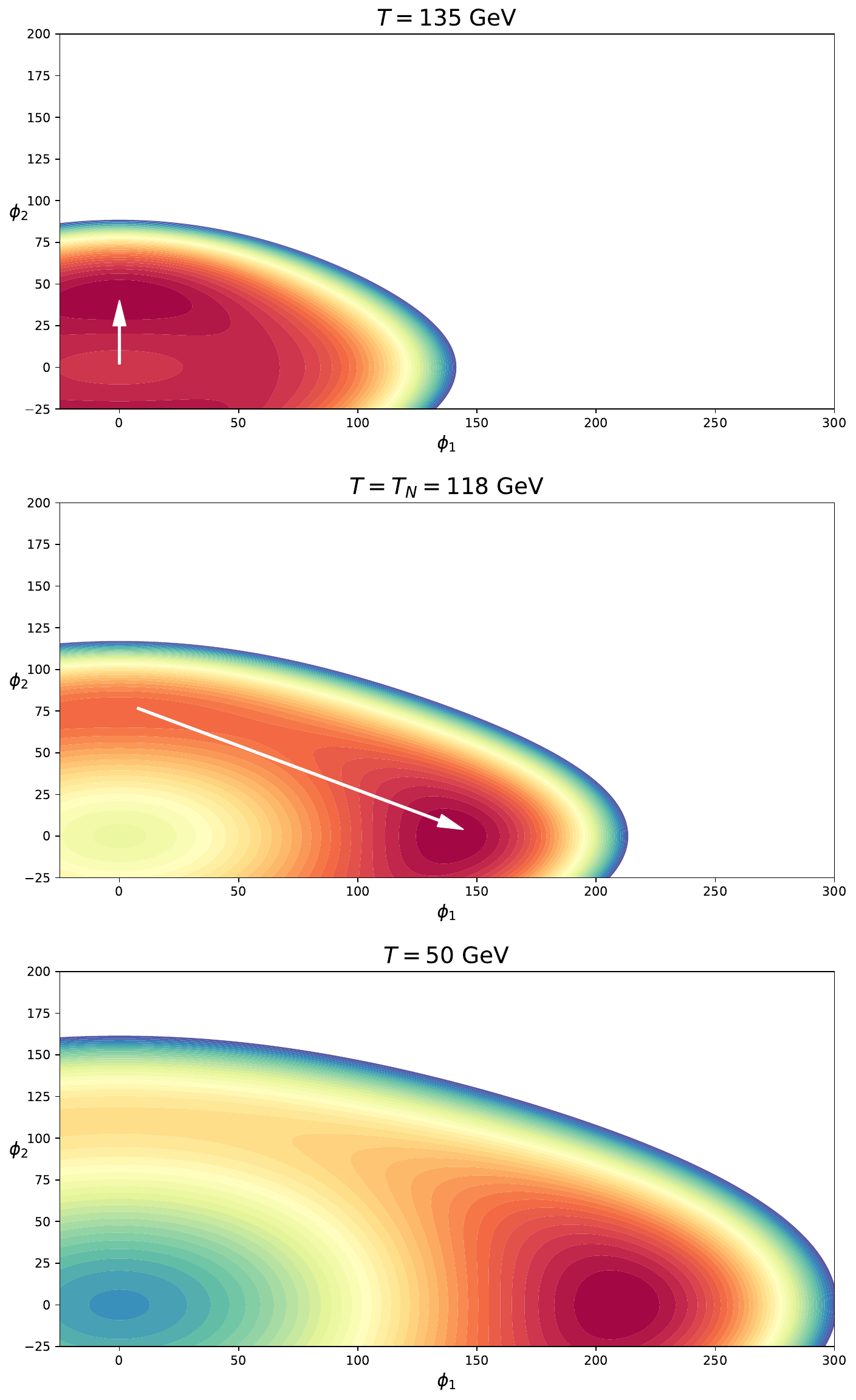}
    \caption{The effective potential of the dimensionally reduced $\mathbb{Z}_2$ scalar singlet model at leading order. Top: the potential calculated at $T = 135\gev$, at which point the scalar minima is formed. The smooth-crossover $(0, 0) \rightarrow (0, s)$ into this minimum is annotated. Middle: the potential at the nucleation temperature $T = 118 \gev$, with an arrow indicating the EWSB transition $(0, s) \rightarrow (h, 0)$. Bottom: the potential at $T = 50\gev$, at which point the scalar minimum has dissolved and the Higgs minimum remains the stable vacuum. }
    \label{fig:3dEFT_plot}
\end{figure}

\begin{figure}
    \centering
    \includegraphics[width=\linewidth]{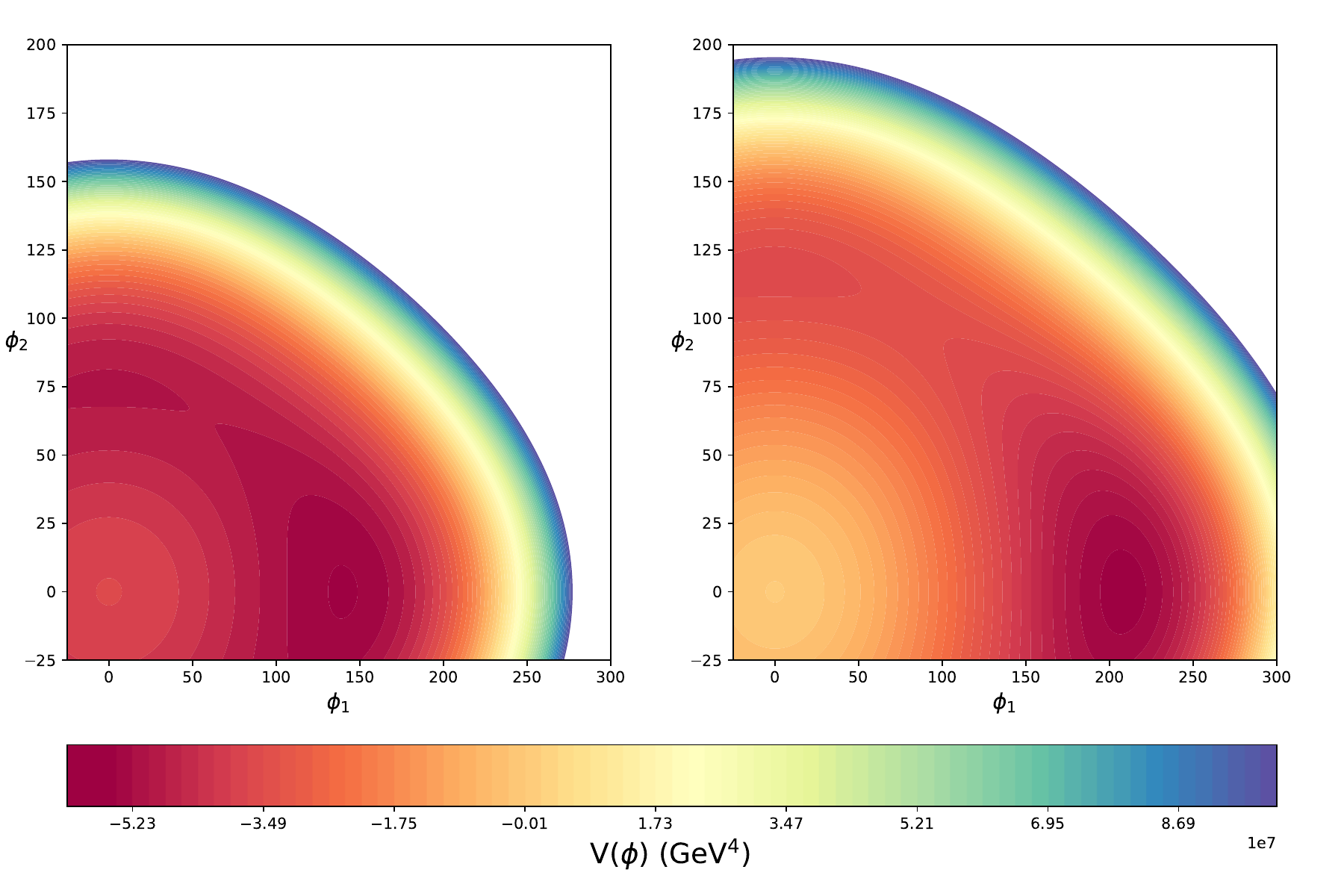}
    \caption{A comparison of the effective potential calculated at the nucleation temperature for the tree-level potential (left), and the one-loop potential (right). On this plot the additional barrier between the respective minima can be seen.}
    \label{fig:3dEFT_NLO_LO}
\end{figure}

The behaviour described above is altered by radiative corrections. In particular, the first step is now of first order, with a barrier being generated between the phases $(0, 0)$ and $(s, 0)$ (although this barrier is negligible from the perspective of gravitational waves). The second step remains first-order, as it was at tree-level. However, the one-loop corrections alter the thermal parameters $\alpha$ and $\beta$ of this second step, resulting in a different gravitational wave spectrum. A summary of the differences between the tree-level and one-loop thermodynamics for the EWSB transition is shown in \cref{tab:3dLOvNLO}. The one-loop potential provides an additional barrier for the $(s,0) \rightarrow (0, h)$ transition, leading to a more strongly supercooled transition, with $\delta_{sc} = 0.06 \ (0.29)$ for the LO (NLO) transition. Consequently the transition has both a larger $\alpha$ and smaller $\beta/H$, both indicators of a stronger gravitational wave signal (c.f \cref{sec:GravWaves}). Indeed, this is seen in a signal approximately 7 orders of magnitude stronger. Additionally, the peak amplitude is now seen at 10\,mHz, near the prospective detectability threshold of LISA. Note however that for this benchmark point, neither signal exceeds the $\text{SNR} > 10$ threshold condition. Lastly, in \cref{fig:3dEFT_NLO_LO} we show each potential at their respective nucleation temperature. On this plot the additional off-diagonal barrier separating the two minima is shown to be more prominent, leading to the behaviour just outlined.

\begin{table*}[t]
\centering
\begin{tabularx}{0.9 \textwidth}{X X X X X X X}
  \bottomrule
  & $T_C$ & $T_N$ & $\alpha$ & $\beta/H$ & Peak $\Omega h^2$ & Peak freq.\\
  \midrule
  Tree-level & 125.388 & 118.444 & 0.00157 & 2950.73 & $3.03 \times 10^{-22}$ & 0.1038 \\
  \midrule
  One-loop & 110.684 & 78.831 &  0.02317 & 407.124 & $4.38\times 10^{-17}$ & 0.0096 \\
  \bottomrule
\end{tabularx}
\caption{Comparison of thermodynamics for the second transition $(0, s) \rightarrow (h, 0)$ across the tree-level and one-loop potentials in the dimensionally reduced scalar singlet model. The stronger supercooling in the one-loop case leads to a much stronger gravitational wave signal. In addition, the peak frequency has been moved much closer to the mHz band, in which LISA is expected to operate.}
\label{tab:3dLOvNLO}
\end{table*}

\section{Conclusions}\label{sec:Conc}

In this work, we present \code{PhaseTracer2}, an advanced \code{C++} software tool designed for modelling of cosmological first-order phase transitions in theories with multiple scalar fields.  This tool represents a significant upgrade over its predecessor, expanding capabilities across both the theoretical modelling and computational realms. The improvements in \code{PhaseTracer2} address critical needs for analysing cosmological phase transitions and the gravitational wave signatures that they might generate, with applications in both particle physics and cosmology.

\code{PhaseTracer2} extends its utility through several key advancements.  It accommodates multiple schemes for evaluating the effective potential, including the \msbar scheme, $R_\xi$ gauge, and various high-temperature expansions, which allow users to explore a broader set of physical scenarios.  \code{PhaseTracer2} also integrates effective potentials derived from the dimensionally reduced models from \code{DRalgo}, thus enabling calculations in three-dimensional effective field theory (3DEFT).  This enhancement allows for accurate calculation of thermal masses and coupling dynamics critical to high-temperature cosmology.

A significant enhancement is the implementation of comprehensive bounce action and nucleation temperature calculations in \code{PhaseTracer2}.  Since the dynamics of first-order phase transitions are governed by bubble nucleation and expansion, \code{PhaseTracer2} implements advanced methods for computing the bounce action.  Through both a native multi-dimensional path deformation approach and an interface to the external code \code{BubbleProfiler}, \code{PhaseTracer2} calculates bubble nucleation rates, essential for determining transition completion.

In response to the growing interest in gravitational wave detection from cosmological sources, \code{PhaseTracer2} includes modules for predicting gravitational wave spectra arising from bubble collisions, sound waves, and turbulence in the early Universe.  These predictions are critical for connecting theoretical models that predict phase transitions, to experimental observation such as LISA, Taiji and TianQin, potentially enabling the identification of gravitational waves from the early Universe.

\code{PhaseTracer2} maintains the fast and stable performance of the original \code{PhaseTracer} while introducing a flexible and modular structure, allowing users to extend its functionalities through custom potential classes and model-specific parameters.  The code structure allows for easy integration of new models and methods, thereby broadening the application scope to various beyond-Standard Model theories.  

In summary, \code{PhaseTracer2} represents a powerful and versatile tool for investigating cosmological phase transitions, with wide-ranging applications in understanding the physics of the early Universe and potentially contributing to gravitational wave astronomy.  When used in combination with \code{TransitionSolver} \cite{Athron:2024tbc}, the bubble dynamics and gravitational wave spectrum are treated as well as possible without simulations. Future developments may include further refinements in renormalization scheme flexibility, improved methods for gauge independent nucleation rates, and additional integration with bubble dynamics simulators and gravitational wave calculators, allowing \pt to remain at the forefront of computational tools in particle cosmology.

\begin{acknowledgement}
C.B. is supported by the Australian Research Council grants DP210101636, DP220100643 and LE21010001.  The work of PA is supported by National Natural Science Foundation of China (NNSFC) Key Projects grant No.~12335005, the supporting fund for foreign experts grant wgxz2022021 and the NNSFC RFIS-II grant No.~12150610460. YZ is supported NNSFC No.~12105248, No.~12335005,     and Henan Postdoctoral Science Foundation No.~HN2024003. AF was supported by RDF-22-02-079. We thank I. Weeden with her assistance in debugging \code{PhaseTracer}.
\end{acknowledgement}
\appendix
\section{Strength of the Phase Transition}
\label{sec:alpha}
To compute the GW spectrum, we need to know the strength of the FOPT. We quantify the strength by $\alpha$, the fraction of available potential energy available for the the production of GWs. We classify transitions according to 
\begin{itemize}
    \item $\alpha \lesssim \mathcal{O}(0.01)$ --- weak transitions
    \item $\alpha \sim \mathcal{O}(0.1)$ --- intermediate transitions
    \item $\alpha \gtrsim \mathcal{O}(1)$ --- strong transitions
\end{itemize}
There are, however, different definitions of $\alpha$ used in the literature.  For the bag model, the pressure density $p_{f}$ ($p_{t}$) and energy density $e_{f}$ ($e_{t}$) in the false vacuum (true vacuum) are simplified as \cite{Espinosa:2010hh} 
\begin{align}
    p_{f} = \frac{1}{3}a_{f} T^4 - \epsilon, ~~e_{f} &= a_{f}T^4 + \epsilon\\
    p_{t} = \frac{1}{3}a_{t} T^4 ,~~ e_{t} &= a_{t}T^4.
\end{align}
This bag model equation of state is a simplified model where  the pressure (or equivalently potential since $p= -V$) and energy density only receive a constant vacuum contribution and radiation contribution that goes like $T^4$.  This is equivalent to assuming that all particles either have masses so large that they can be integrated out or so small that the masses are negligible compared to temperature such that the $T^4$ term in the finite temperature effective potential dominates.  

The $\alpha$ in the bag model, which we denote as $\alpha_{\epsilon}$, is defined as
\begin{equation}
    \alpha_{\epsilon} = \frac{\epsilon}{a_{f}T_{*}^4} = \frac{4 \epsilon}{3 \left[e_{f}(T_{*}) + p_{f}(T_{*})\right]}=\frac{4 \epsilon}{3 w_{f}(T_{*})},
\end{equation}
where $w_{f}(T_{*})$ is the enthalpy density far away from the bubble wall and $T_{*}$ is the reference temperature. 

For the bag model equation of state, Ref.~\cite{Espinosa:2010hh} determines relations between the $\alpha$, the bubble wall velocity and an efficiency coefficient $\kappa$ which is also required in order to compute the kinetic energy that contributes to the gravitational wave signals.  However since in realistic models the pressure and energy density do not neatly separate into constant vacuum and $T^4$ radiation density one must use a different definition of $\alpha$.  There are several different definitions used in the literature, but replacing the constant vacuum energy with the trace anomaly  $\theta = \frac{1}{4} (e - 3p)$ is most appropriate since $e = \theta + \frac{3}{4} w$, with $\theta$ representing the part of the energy density that is liberated by the vacuum, rather than immediately contributing to heat ($e_Q = \frac{3}{4} w$),  
\begin{equation} \label{eq:alpha_definition}
    \alpha_{\theta} \equiv \frac{4 (\theta_f - \theta_t)}{3 w_{f}}, 
\end{equation}
where $\theta_f$ and $\theta_t$ are the trace anomaly evaluated in the false and true vacuums respectively. We can also approximate the enthalpy density in the denominator by replacing it with the radiation energy density,
\begin{equation}
    \alpha_{\theta} \approx \frac{(\theta_f - \theta_t)}{\pi^2 g_* T_{*}^4/30} = \frac{\left(V(\phi) - \frac{T}{4} \frac{\partial V(\phi, T)}{\partial T}\right) \bigg |^{\phi_f}_{\phi_t}}{\pi^2 g_* T_{*}^4/30},
    \label{eq:alpha_theta_approx}
\end{equation}
where $\phi_f$ and $\phi_t$ represent the vacuum expectation value of the false vacuum and true vacuum, respectively and in the numerator on the right hand side we have expressed the trace anomaly in terms of the potential using the expressions for the energy density $e$ and the pressure $p$.  Making this approximation in place of \cref{eq:alpha_definition} has been found to be accurate to within $1\%$ in the resulting gravitational wave signals when studied within the context of a scalar singlet extension of the Standard Model \cite{Athron:2023rfq}.   

In \code{PhaseTracer} we use the trace anomaly definition of $\alpha$, using the approximation \cref{eq:alpha_theta_approx}.  In the main text section, $\alpha$ always refers to the trace anomaly definition written as $\alpha_\theta$ here.

Departures from bag model assumptions to allow the speed of sound to vary away from its bag model value have been captured in Refs.~\cite{Giese:2020znk,Giese:2020rtr}.  These effects are not included in \pt, but are included in the \code{TransitionSolver} software \cite{Athron:2024tbc} where many other subtle effects are accounted for. 

\newpage
\bibliographystyle{JHEP}
\bibliography{bibliography}
\end{document}